\documentclass[aps,prl,twocolumn,groupedaddress]{revtex4-2}

\usepackage{graphicx}
\usepackage{epstopdf}
\usepackage{amssymb}
\usepackage{amsmath}
\usepackage{color}
\usepackage{hyperref}
\usepackage{tikz}
\usepackage{mathrsfs}
\usepackage{wasysym}
\usepackage{tikz-cd}
\usepackage{adjustbox}
\usepackage{booktabs}
\usepackage{orcidlink}
\usepackage{hyperref}

\usepackage{alphalph}
\renewcommand
\thesubsection{\mbox{\AlphAlph{\value{subsection}}}}

\DeclareMathOperator*{\argmin}{arg\,min}
\newcommand{\avrg}[1]{\left\langle #1 \right\rangle}

\newcommand{\krn}{\mathrm{krn}}
\newcommand{\rng}{\mathrm{rng}}
\newcommand{\rnk}{\mathrm{rnk}}
\newcommand{\nll}{\mathrm{nll}}
\newcommand{\grad}{\mathrm{grad}}
\newcommand{\curl}{\mathrm{curl}}
\newcommand{\dive}{\mathrm{div}}

\newcommand{\RR}{\mathbb{R}}
\newcommand{\CC}{\mathbb{C}}

\newcommand{\FF}{\mathbb{F}}

\begin{document}


\title{
Analysis of the inference of ratings and rankings in complex networks using discrete exterior calculus on higher--order networks
}


\author{Juan I. Perotti\,\orcidlink{https://orcid.org/0000-0001-7424-9552}}
\email[]{juan.perotti@unc.edu.ar}
\affiliation{
{\small
Instituto de F\'isica Enrique Gaviola (IFEG-CONICET),\\
Facultad de Matem\'atica, Astronom\'ia, F\'isica y Computaci\'on, Universidad Nacional de C\'ordoba,\\
Ciudad Universitaria, 5000 C\'ordoba, Argentina
}
}


\date{\today}

\begin{abstract}
The inference of rankings plays a central role in the theory of social choice, which seeks to establish preferences from collectively generated data, such as pairwise comparisons.
Examples include political elections, ranking athletes based on competition results, ordering web pages in search engines using hyperlink networks, and generating recommendations in online stores based on user behavior. 
Various methods have been developed to infer rankings from incomplete or conflicting data. One such method, HodgeRank, introduced by Jiang {\em et al.}~[Math. Program. {\bf 127}, 203 (2011)],
utilizes Hodge decomposition of cochains in higher--order networks to disentangle gradient and cyclical components contributing to rating scores, enabling a parsimonious inference of ratings and rankings for lists of items.
This paper presents a systematic study of HodgeRank's performance under the influence of quenched disorder and across networks with complex topologies generated by four different network models. 
The results reveal a transition from a regime of perfect retrieval of true rankings to one of imperfect retrieval as the strength of the quenched disorder increases. 
A range of observables are analyzed, and their scaling behavior with respect to the network model parameters is characterized. 
This work advances the understanding of social choice theory and the inference of ratings and rankings within complex network structures.
\end{abstract}


\maketitle

\section{Introduction}

The rating and ranking candidates, agents, items or options based on collectively expressed preferences is deeply ingrained modern human civilization, making the study of inference methods for ratings and rankings under diverse conditions highly significant.
Ratings and rankings are often derived from pairwise comparisons, which naturally form a network structure. 
Understanding how the topology of these networks influences the accuracy of rating and ranking inference is of critical importance.

The modern theory of ratings and rankings has its origins in the field of social choice, with early foundations laid in the late 13th century by Ramon Llull in his {\em Ars Electionis}~\cite{colomer2013ramon}. 
Llull's work introduced the concept of pairwise comparisons, the majority principle, and fairness in voting systems. 
Unfortunately, his contributions were largely forgotten over time. 
A mathematical revival of these ideas emerged in the late 18th century with the works of Borda and Condorcet~\cite{borda1781memoire,condorcet1785essai}, who laid the groundwork for modern voting and ranking theory. 
Later, in 1929, Zermelo proposed a model for ranking inference using ratings as latent variables within a probabilistic framework~\cite{zermelo1929berechnung}, but his work was also overlooked~\cite{newman2023efficient}. 
In the 1950s, Bradley and Terry reintroduced Zermelo’s model in a broader context~\cite{bradley1952rank}. 
In the 1960s, a model based on similar principles was introduced and promoted by \'El\H{o}, eventually becoming the default rating system used by the World Chess Federation~\cite{elo1978rating}.
More recently, the advent of the World Wide Web and search engines highlighted the significance of network topology in ranking systems, exemplified by the success of PageRank, the original algorithm behind Google and introduced by Brin and Page at the end of the 90's~\cite{brin1998anatomy}. 
The study of rating and ranking systems remains an active area of research~\cite{jiang2011statistical,csato2017ranking,debacco2018physical,newman2022ranking,jerdee2023luck}, as collective decision-making plays a vital role in modern civilization, particularly in the information age.

In 1736, Euler solved the Seven Bridges of K\"onigsberg problem, establishing the foundations of graph theory and foreshadowing the development of topology. 
Graphs, or networks, can be viewed as a specific type of topological space. 
Within the theory of higher--order networks~\cite{bianconi2021higher,battiston2021physics,majhi2022dynamics,bick2023what,ferraz2024contagion,muolo2024global,bianconi2024nature,bianconi2024theory}, hypergraphs extend the concept of graphs by incorporating hyperlinks, that may be used to represent many-body interactions or multiary relations.
However, unlike graphs, hypergraphs are not necessarily topological spaces, as they may not satisfy the condition of topological closure. 
For example, a hypergraph may include a hyperlink $\{i,j,k\}$ without containing the link $\{i,j\}$. 
When hypergraphs satisfy the closure condition, they form a special class of topological spaces known as abstract simplicial complexes, or simply, simplicial complexes.

Although terminology and definitions may vary by context, many concepts from differential geometry on smooth manifolds can be adapted to the discrete or combinatorial setting of simplicial complexes~\cite{bobenko2008discrete,grady2010discrete,aygun2011spectral,tuncer2015spectral,rovelli2019natural,bianconi2021higher}.
Specifically, within the framework of discrete exterior calculus, discrete or combinatorial analogs of differential forms, gradients, divergences, and Laplacians can be defined on simplicial complexes.
This is the reason for which simplicial complexes are often preferred over general hypergraphs.
In particular, and thanks to the property of topological closure, combinatorial Hodge theory can be applied to decompose cochains on simplicial complexes into exact, coexact and harmonic components, much like the analog decomposition of tensor fields~\cite{grady2010discrete,lim2020hodge}.
This is the approach followed by Jian et al.~\cite{jiang2011statistical} who developed HodgeRank, a method for the inference of ratings that leverages Hodge theory over simplicial complexes to disentangle different contributions to the inferred ratings.
The present article takes advantage of HodgeRank to study the inference of ratings and rankings in complex topologies~\cite{newman2010networks,perotti2009emergent}.
Section~\ref{sec:theory} provides an introduction to the theory of rating and ranking inference, and focuses on HodgeRank. Section~\ref{sec:results} presents the results of numerical experiments using HodgeRank applied to networks with complex topologies. 
Section~\ref{sec:discussion} discuss the results.
Section~\ref{sec:conclusions} summarizes the findings and outlines directions for future research. 

\section{
Theory
\label{sec:theory}
}

Several methods exist for inferring ratings and rankings from pairwise comparison data~\cite{jiang2011statistical}. This section focuses on a particular approach inspired by the foundational works of Zermelo~\cite{zermelo1929berechnung} and Bradley and Terry~\cite{bradley1952rank}, to illustrate how pairwise comparisons can be associated with probabilities that, in turn, relate to rating scores. These concepts are then used to introduce the HodgeRank method, followed by a discussion of the numerical challenges that arise within this framework.

Consider a set of $N=n+1$ items indexed by $i \in \{0, 1, \dots, n\}$. 
These items or agents could represent movies subject to pairwise comparison or even chess players competing in chess tournaments~\cite{perotti2013innovation}. 
For any two items $i$ and $j$, there exists a (possibly empty) sequence $s_{ij} = (s_{ij1}, s_{ij2}, \dots, s_{ijt}, \dots)$ of comparison results, where $s_{ijt} = -s_{jit}$ takes the value $1$, $0$, or $-1$ depending on whether item $i$ wins, draws, or loses the $t$th comparison against item $j$, respectively. 
For example, these results could represent outcomes of chess games between players, or voting preferences over movies. 
Importantly, the set of nonempty sequences, or {\em pairings}, defines the links of an undirected network with $N$ nodes and no self-links, described by an adjacency matrix $a$ with entries $a_{ij} = a_{ji} \in \{0, 1\}$ for $i,j \in \{0, 1, \dots, n\}$. 
The degree, or the number of links adjacent to node $i$, is $k_i = \sum_j a_{ij} = \sum_j a_{ji}$, and the average degree is $\bar{k} = N^{-1} \sum_i k_i = 2M/N$, where $M = \frac{1}{2} \sum_{ij} a_{ij}$ is the total number of links.

Assuming that the results within a sequence are statistically independent and are identically distributed, the probability for a particular sequence to occur is 
\begin{eqnarray}
\label{eq:1}
P(s_{ij1},\hdots,s_{ijt},\hdots)
&=&
P(s_{ij1})\hdots P(s_{ijt})\hdots
\\
&=&
p_{ij}^{x_{ij}}q_{ij}^{y_{ij}}r_{ij}^{z_{ij}}
\nonumber
\\
&=&
v_{ij}^{x_{ij}}(1-v_{ij})^{z_{ij}}
q_{ij}^{y_{ij}}(1-q_{ij})^{x_{ij}+z_{ij}}
,
\nonumber
\end{eqnarray}
where $p_{ij}:=P(S_{ijt}=1)$, $q_{ij}:=P(S_{ijt}=0)$ and $r_{ij}:=P(S_{ijt}=-1)$, and $x_{ij}=\sum_t \delta_{s_{ijt},1}$, $y_{ij}=\sum_t \delta_{s_{ijt},0}$ and $z_{ij}=\sum_t \delta_{s_{ijt},-1}$ are the number of wins, draws and losses of $i$ over $j$ for all $t$, respectively.
Moreover, $v_{ij}:=P(S_{ijt}=1|S_{ijt}\neq 0)$ and $1-v_{ij}:=P(S_{ijt}=-1|S_{ijt}\neq 0)$ since $p_{ij}=v_{ij}(1-q_{ij})$ and $r_{ij}=(1-v_{ij})(1-q_{ij})$.
There are different ways to fit the probabilities of Eq.~\ref{eq:1} to given data.
For instance, in the maximum {\em a posteriori} bayesian approach of the multinomial distribution with a uniform prior, the expected value of the probabilities are $p_{ij}=(x_{ij}+1)/(x_{ij}+y_{ij}+z_{ij}+3)$, $q_{ij}=(y_{ij}+1)/(x_{ij}+y_{ij}+z_{ij}+3)$ and $r_{ij}=(z_{ij}+1)/(x_{ij}+y_{ij}+z_{ij}+3)$, and therefore $v_{ij}=p_{ij}/(1-q_{ij})=(x_{ij}+1)/(x_{ij}+z_{ij}+2)$.
In this way, pairs of items with few comparisons can be handled appropriately.
In particular, if the kind of comparisons under consideration admit no draws, then $q_{ij}=0$ for all $ij$.

\subsection{
A rating method from least squares
\label{sec:ratings_and_rankings_lsqr}
}

Consider the parametrization of the probabilities $v_{ij}$ in Eq.~\ref{eq:1} using the logistic function $v = (1 + e^f)^{-1}$. 
This defines a monotonically decreasing bijection between the probabilities $v_{ij} \in (0, 1)$ and the corresponding values $f_{ij} = \ln(v_{ij}^{-1} - 1) \in (-\infty, \infty)$, and motivates the following question.
Is there a vector $w \in \mathbb{R}^N$ with components $w_i$ such that Eq.~\ref{eq:2} is satisfied?
\begin{equation}
\label{eq:2}
f_{ij}=w_j-w_i
.
\end{equation}
If such a vector exists, it could serve as a rating system, where $v_{ij} > 1/2$ if and only if $f_{ij} < 0$, or equivalently, if and only if $w_j < w_i$. 
Additionally, this vector could be used to derive a ranking system, where a permutation $r_0, \dots, r_n$ of $0, 1, \dots, n$ satisfying $w_{r_0} \leq w_{r_1} \leq \dots \leq w_{r_n}$, yields a ranking $r_0, r_1, \dots, r_n$ of the items.
However, such a vector $w$ does not always exist. 
For example, a dataset might exhibit cyclic behavior, where item $i$ defeats $j$, $j$ defeats $k$, and $k$ defeats $i$, leading to inconsistencies in the ranking. 
Cyclical inconsistencies are common in practice.
Fortunately, a weakened version of Eq.~\ref{eq:2} can still be used to infer useful ratings and rankings. 
Specifically, the exact condition of Eq.~\ref{eq:2} can be replaced by the least-squares problem in Eq.~\ref{eq:3}~\cite{csato2015graph}.
\begin{equation}
\label{eq:3}
w = \argmin_{w'}
\sum_{ij} a_{ij}\,|f_{ij} - (w'_j-w'_i)|^2
.
\end{equation}
The idea here is that, although Eq.~\ref{eq:2} cannot always be satisfied, the solution to Eq.~\ref{eq:3} provides the vector $w$ that best approximates the intended behavior in a least-squares sense.

\subsection{HodgeRank}

In this section the HodgeRank method is reviewed~\cite{jiang2011statistical}.
For further reading, some practical applications of HodgeRank are demonstrated by Johnson and Goldring~\cite{johnson2013discrete} and some numerical challenges emerging from its appliation to complex networks are studied by Hirani {\em et al.}~\cite{hirani2015graph}.
Moreover, an elementary review about the discrete exterior calculus and combinatorial Hodge theory on simplicial complexes is provided in the Appendix, where the formal details are introduced.

A hypergraph is composed of nodes, links and higher--order links or hyperlinks such as triangles, tetrahedra and so on.
A simplicial complex is a specific type of hypergraph that satisfies the property of topological closure, making it a topological space. 
This structure allows for the definition of cochains, which are discrete or combinatorial versions of differential forms where scalar forms correspond to 0-cochains, vector forms correspond to 1-cochains and, more generally, $k$-forms correspond to $k$-cochains.
In practical terms, a 0-cochain is a vector of scalars $w_i$ associated with nodes $i$, while a 1-cochain is a vector of antisymmetric scalars $f_{ij} = -f_{ji}$ associated with links $ij$. 
More generally, a $k$-cochain is a vector of alternating scalars $t_{i_0 \dots i_k}$ associated with hyperlinks known as $k$-simplices. 
Crucially, discrete or combinatorial versions of differential operators such as the gradient, the divergence, and the Laplacian, can be defined on simplicial complexes.
This framework allows the least-squares problem in Eq.~\ref{eq:3} to be reformulated in terms of the 3-clique complex $K = K_0 \cup K_1 \cup K_2$, derived from the network of adjacency matrix $a$. 
Here, the values $f_{ij}$ are recognized as the components of a 1-cochain $f$, where $f_{ij} = -f_{ji}$ because $v_{ji} = 1 - v_{ij}$, while the differences $w_j - w_i$ correspond to the components of the gradient of a 0-cochain $w$.
More specifically, Eq.~\ref{eq:3} can be rewritten as
\begin{equation}
\label{eq:4}
w=\argmin_{w'\in C_0^*}|f-d_0(w')|^2
,
\end{equation}
where $w \in C_0^*$ is a 0-cochain, and $d_0 = \grad$ is the gradient operator on the vector space $C_0^*$ of 0-cochains.
Thanks to this identification, the least-squares problem in Eq.~\ref{eq:4} can be reformulated in terms of the Hodge decomposition of $f$. 

\begin{figure}
\includegraphics*[scale=.9]{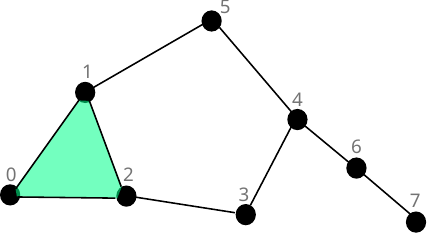}
\caption{
\label{fig:1}
(color online).
An elementary example of a simplicial complex with $N=8$ nodes.
Its set of 0-simplices or nodes is $K_0=\{\{0\},\{1\},...,\{7\}\}$, of 1-simplices or links is $K_1=\{\{0,1\},\{1,2\},\{0,2\},\{2,3\},\{3,4\},\{4,5\},\{1,5\},\{4,6\},\{6,7\}\}$ and of 2-simplices or triangles is $K_2=\{\{0,1,2\}\}$.
A local cyclic flow can only exist in the 2-simplex $\{0,1,2\}$ (depicted in green), while nonlocal cyclic flows can exist in the closed paths $1\to 2\to 3\to 4\to 5\to 1$ and $0\to 2\to 3\to 4\to 5\to 1\to 0$.
}
\end{figure}

In the Hodge decomposition of a general $k$-cochain $f \in C_k^*$, there exist unique $k$-cochains $s, g, h \in C_k^*$ such that $f = s \oplus h \oplus g$, where $s\in \rng\,d_{k}^*$, $h\in \krn\,L_k$ and $g\in \rng\,d_{k-1}$ are called the coexact, the harmonic and the exact components of $f$.
Here, $\rng\,d_{k}^*$, $\krn\,L_k$ and $\rng\,d_{k-1}$ are the ranges and kernel of the $k$th dual coboundary operator $d_k^*$, the $k$th Laplacian $L_k$ and the $(k-1)$th coboundary operator $d_{k-1}$, respectively, and $C_k^*$ is the vector space of $k$-cochains.

To compute the Hodge decomposition of a $k$-cochain $f$, recall that
$\krn\,d_k=\krn\,L_k\oplus\,\rng\,d_{k-1}$
and 
$\krn\,d_{k-1}^*=\rng\,d_k^*\oplus\, \krn\,L_k$.
In consequence
\begin{eqnarray}
\label{eq:5}
d_{k}(f) &=& d_{k}(s + h + g) \\
&=& d_{k}(s) \nonumber \\
&=& d_{k}[d_{k}^*(u)] \nonumber \\
&=& L_{k+1}^{\downarrow}(u) \nonumber \\
\alpha_{k+1} f &=& \alpha_{k+1} \alpha_{k+1}^T u 
\nonumber
\end{eqnarray}
and
\begin{eqnarray}
\label{eq:6}
d_{k-1}^*(f) &=& d_{k-1}^*(s + h + g) \\
&=& d_{k-1}^*(g) \nonumber \\
&=& d_{k-1}^*[d_{k-1}(w)] \nonumber \\
&=& L_{k-1}^{\uparrow}(w) \nonumber \\
\alpha_{k}^T f &=& \alpha_{k}^T \alpha_{k} w 
\nonumber
\end{eqnarray}
are singular linear equations for some $u \in C_{k+1}^*$ and $w \in C_{k-1}^*$, that can be solved to obtain $s = \alpha_{k+1}^T u$, $g = \alpha_{k} w$, and $h = f - s - g$.
Although the Hodge Laplacians 
$L_{k+1}^{\downarrow}$
and
$L_{k-1}^{\uparrow}$
are symmetric operators, they are also singular and, therefore, Eqs.~\ref{eq:5} and \ref{eq:6} still pose numerical challenges. 
Fortunately, these are the normal equations of respective least-squares problems. 
Hence, their singularities can be addressed by seeking the solutions $u$ and $w$ that also minimize the squared norms $|u|^2$ and $|w|^2$, respectively.

The Hodge decomposition provides a solution to Eq.~\ref{eq:4} when $k=1$.
Namely, for $k=1$, any solution $w$ of Eq.~\ref{eq:6} is also a solution to Eq.~\ref{eq:4}, and vice versa. 
This approach to inferring $w$ from the Hodge decomposition of $f$ is what Jiang {\em et al.}~\cite{jiang2011statistical} referred to as {\em HodgeRank}. 
Additionally, drawing an analogy from physics in three dimensions, the coexact, harmonic, and exact components $s$, $h$, and $g$, are referred to as the solenoidal, harmonic, and gradient components of $f$, respectively, providing insightful observations in the context of rating inference.
Namely, the gradient component $g$ represents a noncyclic flow, which aligns perfectly with a rating vector $w$. 
In contrast, the solenoid and harmonic components, $s$ and $h$, correspond to local and nonlocal cyclic flows, respectively, which are incompatible with a consistent rating system (see Fig.~\ref{fig:1} for an example).
These cyclic flows were previously discussed in Sec.~\ref{sec:ratings_and_rankings_lsqr}. 
The norms $|s|$, $|h|$, and $|g|$ help to gauge the extent to which the inferred ratings are affected by local or nonlocal cyclic flows.

\begin{figure*}
\includegraphics*[scale=.4]{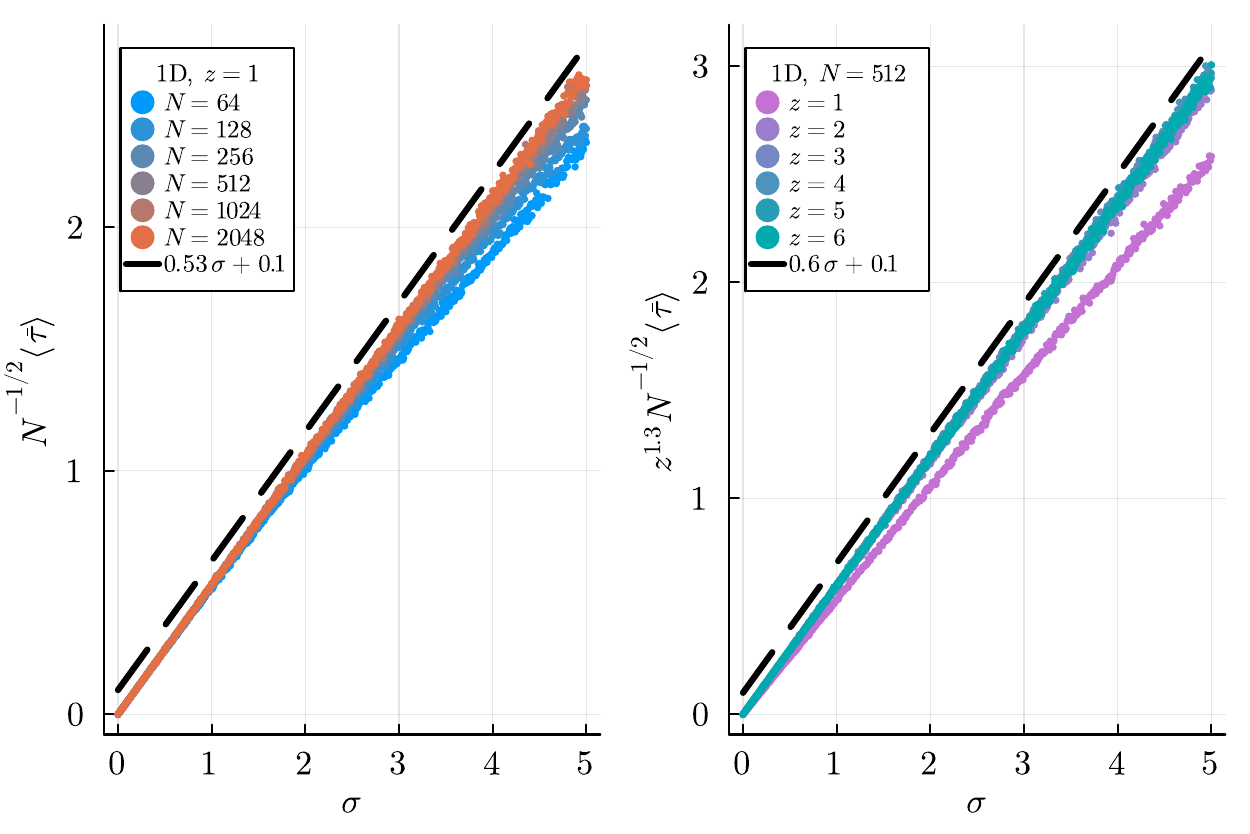}
\includegraphics*[scale=.4]{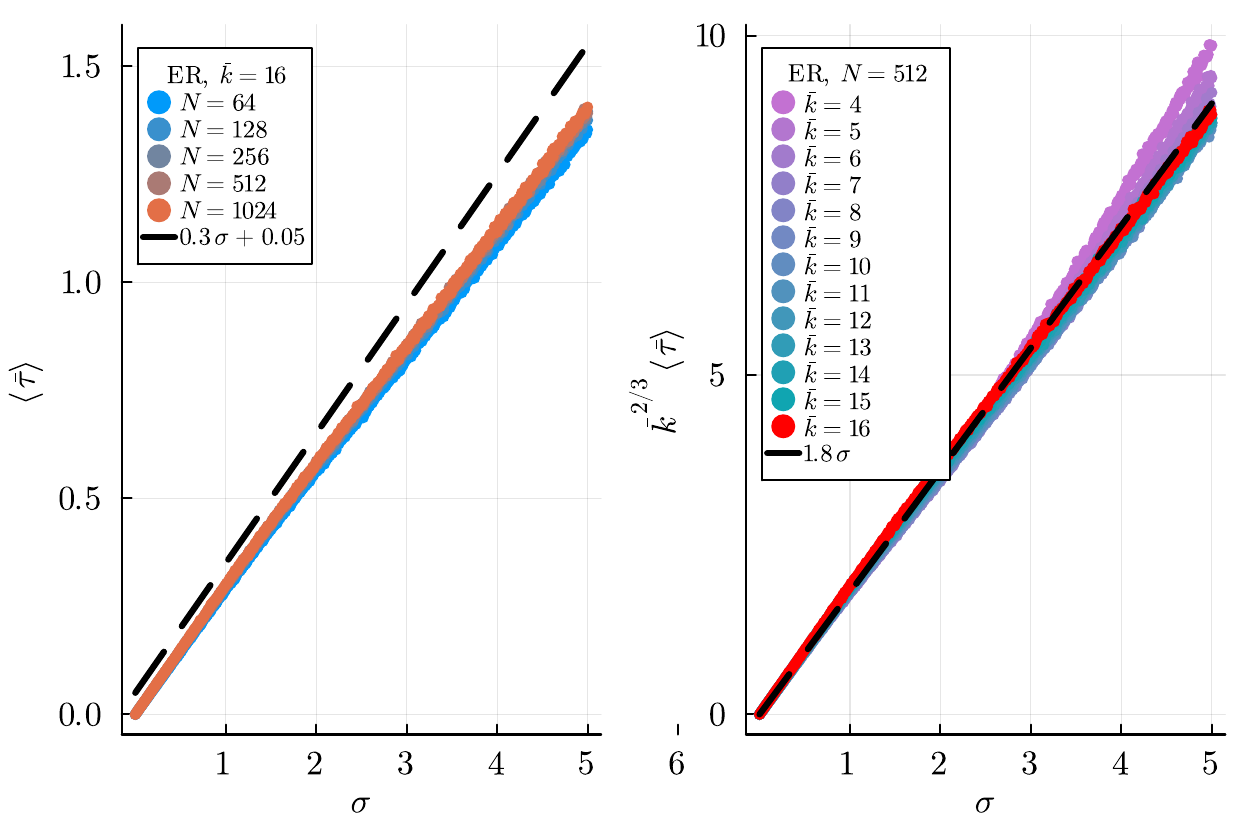}
\put(-483,140){\bf a)}
\put(-240,140){\bf b)}
\\
\includegraphics*[scale=.4]{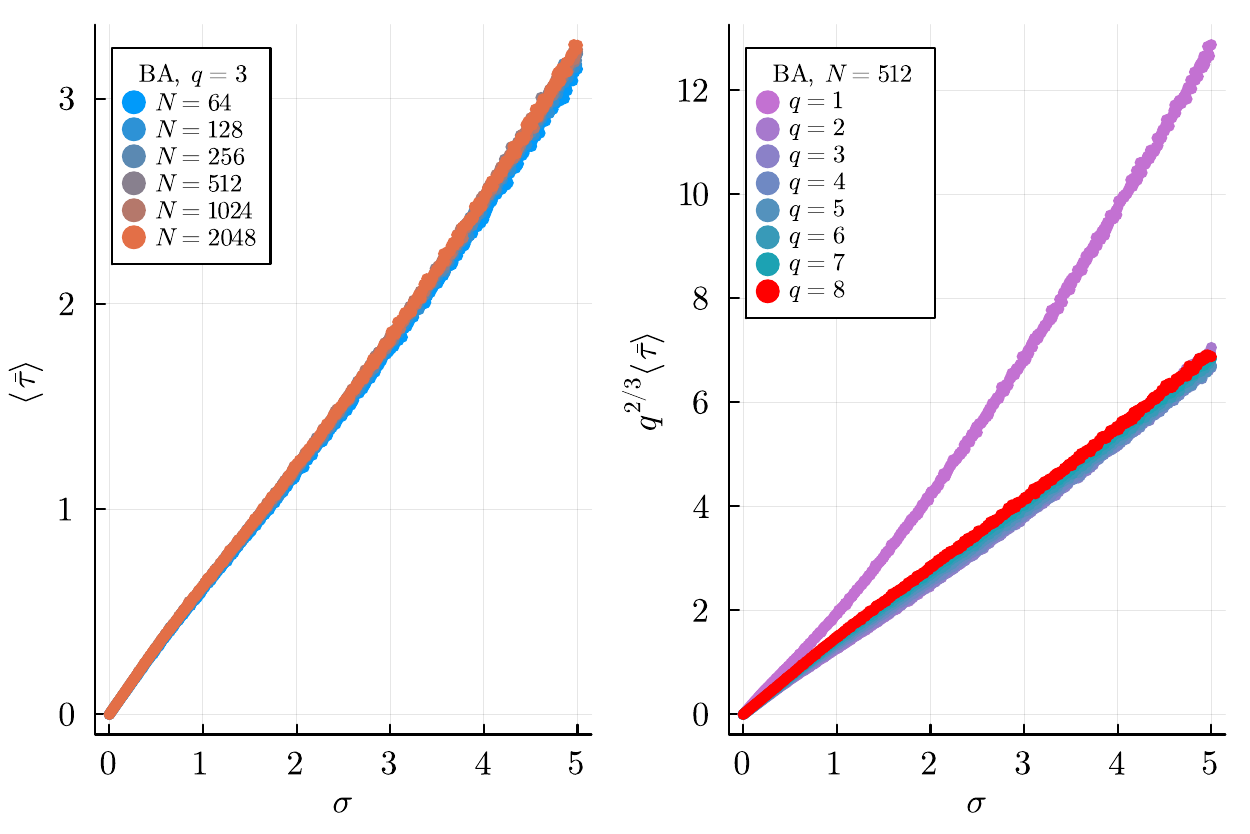}
\includegraphics*[scale=.4]{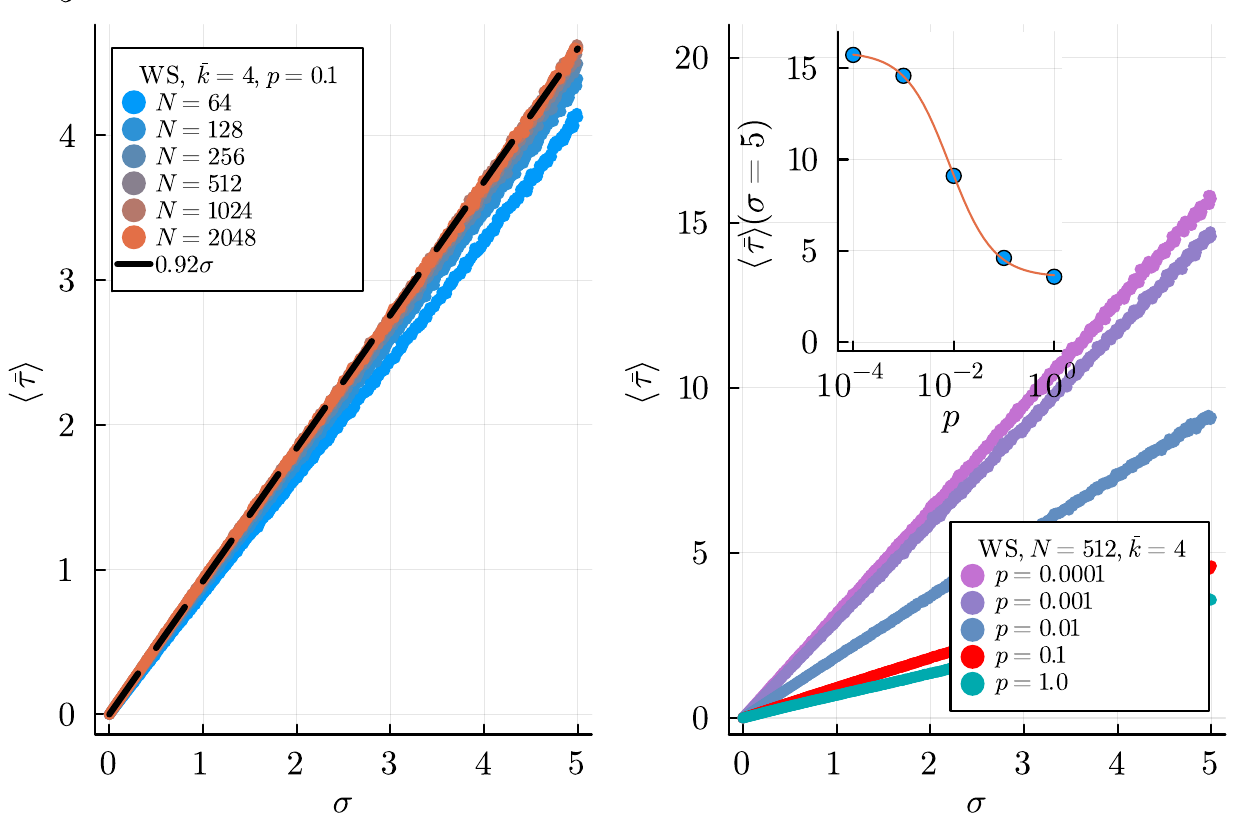}
\put(-483,140){\bf c)}
\put(-240,140){\bf d)}
\caption{
\label{fig:2}
(color online).
The mean of the average of the absolute differences of ratings $\avrg{\bar{\tau}}$ averaged over 5000 network samples and realizations of the disorder $\eta$ as a function of $\sigma$ for different sizes $N$ (left subpanels) and values of the parameter $\theta$ (right subpanels) of the corresponding network model.
Dashed lines represent guides to the eye.
In panel (a), 1D lattices where $\theta=z$ is the number of neighbors per node.
In panel (b), Erd\H{o}s-R\'enyi random networks where $\theta=\bar{k}$ is the average degree.
In panel (c), Barab\'asi-Albert scale-free networks where $\theta=q$ is the number of new links for each new node of the growing algorithm.
In panel (d), Watts-Strogatz small-world networks of average degree $\bar{k}=4$ where $\theta=p$ is randomization probability per link.
In the inset, the value of $\avrg{\bar{\tau}}$ for fixed $\sigma$ is plotted as a function of $p$.
The orange line fits the shifted sigmoid function $\avrg{\bar{\tau}}(\sigma=5)=(15.9\pm 0.2)-(12.4\pm 0.3)/(1+(0.008\pm 0.0004)/p)$ in linear-log scale.
}
\end{figure*}

In practice, some components $w_i$ of the minimum norm solution to Eq.~\ref{eq:6} may be negative. 
Since a rating system with non-negative ratings is often desirable, the components can be shifted by a constant term, $w_i \to w_i - \min_j w_j$. 
This shifted rating vector remains a valid solution to Eq.~\ref{eq:4} and facilitates the comparison between the inferred $w$ and the true ratings $\hat{w}$ when the later are available.

\section{
Results
\label{sec:results}
}

In the following numerical experiments, HodgeRank is applied to infer the ratings and rankings of items matched by networks of various topologies: one-dimensional (1D) lattices, Erdős-Rényi (ER) random graphs, Barabási-Albert (BA) scale-free networks, and Watts-Strogatz (WS) small-world networks~\cite{newman2010networks}.

For each network of adjacency matrix $a$, a vector of true ratings $\hat{w}$ is generated according to $\hat{w}_i = i$ for $i = 0, 1, \dots, n$. 
With this choice, the expected true ranking is $\hat{r}_i=i$.
In 1D lattices and WS networks, nodes are enumerated following the natural order in the lattice and in BA networks nodes are enumerated from the oldest to the newest.
Using the true rating vector $\hat{w}$, a corresponding 1-cochain $f$ is defined by
\begin{equation}
\label{eq:f_and_eta}
f_{ij}
=
\hat{w}_j-\hat{w}_i
+
\eta_{ij},
\end{equation}
where the different components of the 1-cochain $\eta$ are independent and identically distributed random variables drawn from a Gaussian distribution with zero mean and standard deviation $\sigma$. 
Equation~\ref{eq:f_and_eta} is similar to Eq.~\ref{eq:2}, but includes the contribution of the additive quenched disorder characterized by $\eta$ that models the effect of statistical fluctuations and the cyclic flows captured by the components $s$ and $h$ of $f$.
When $\sigma = 0$, the inferred ratings are expected to be perfectly aligned with the true ratings, but as $\sigma$ increases, discrepancies are likely to emerge.

\begin{figure*}
\includegraphics*[scale=.4]{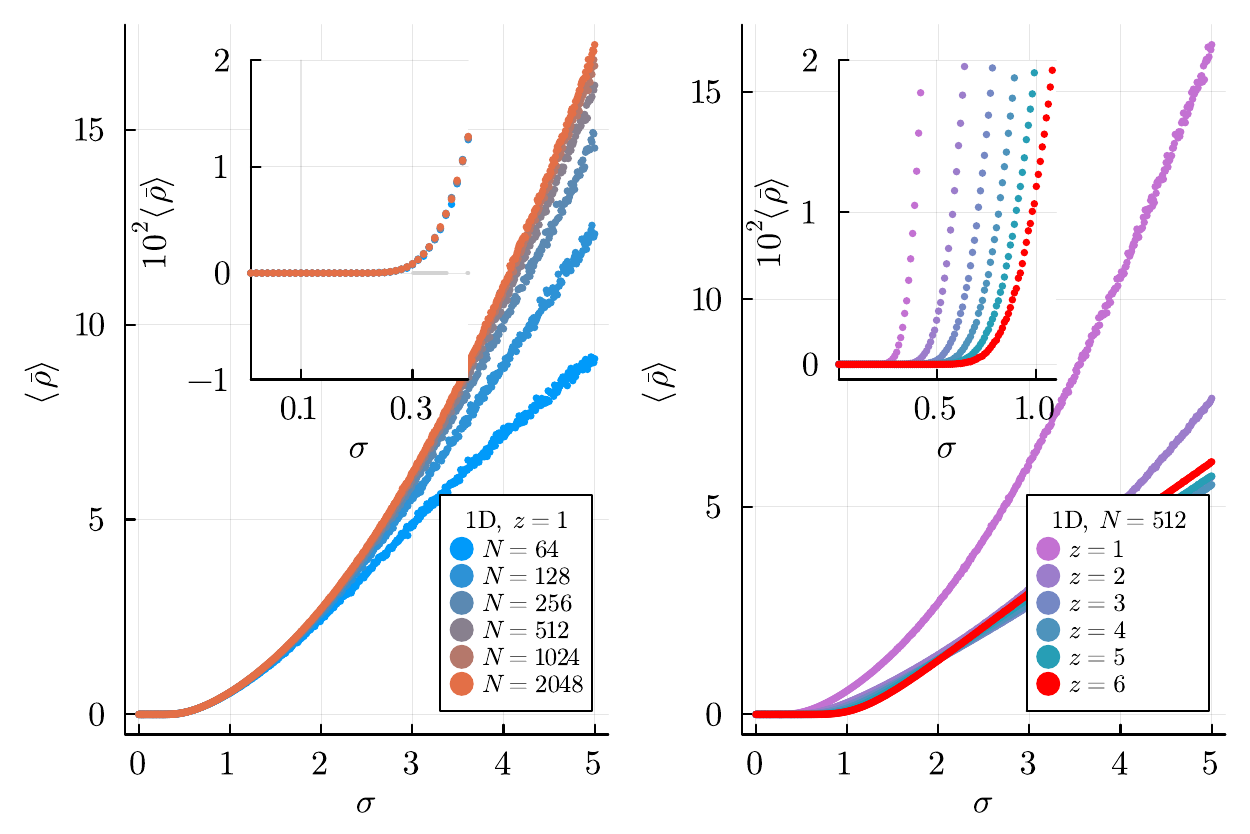}
\includegraphics*[scale=.4]{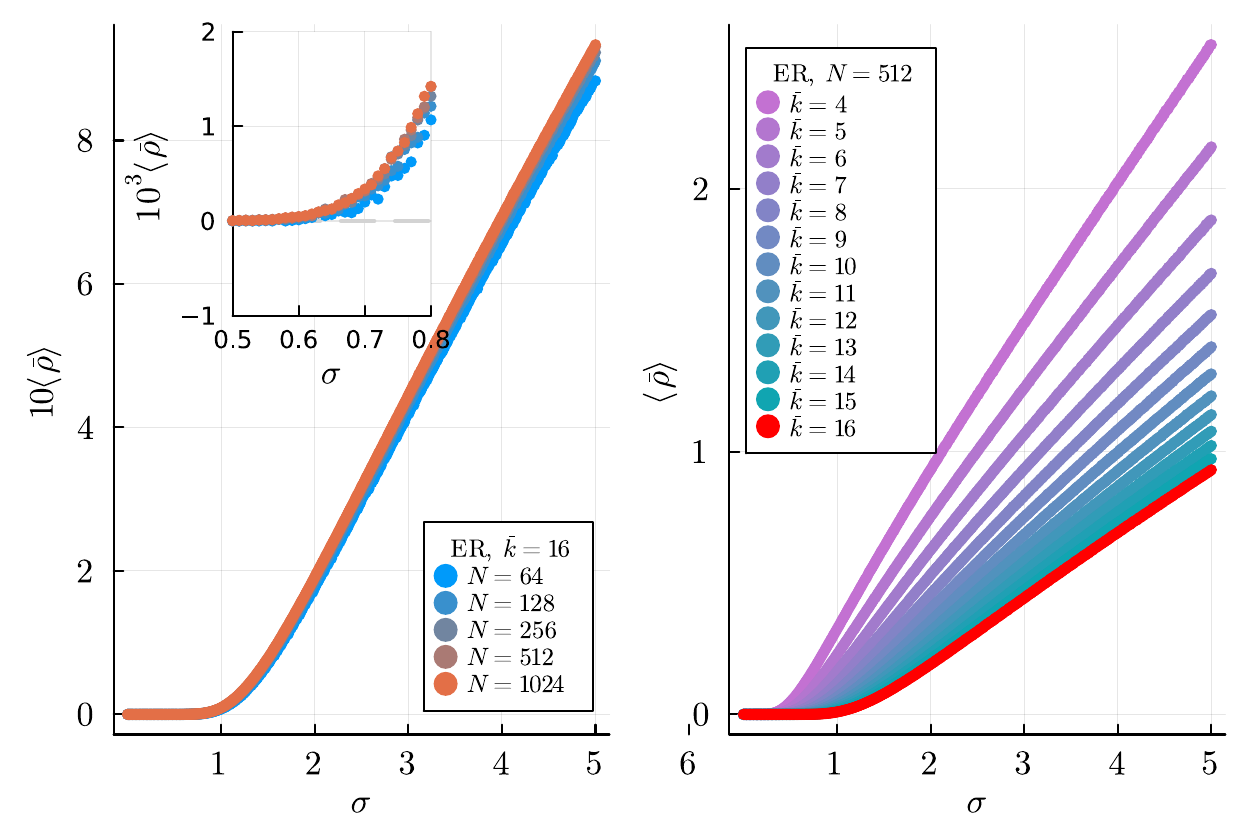}
\put(-483,140){\bf a)}
\put(-240,140){\bf b)}
\\
\includegraphics*[scale=.4]{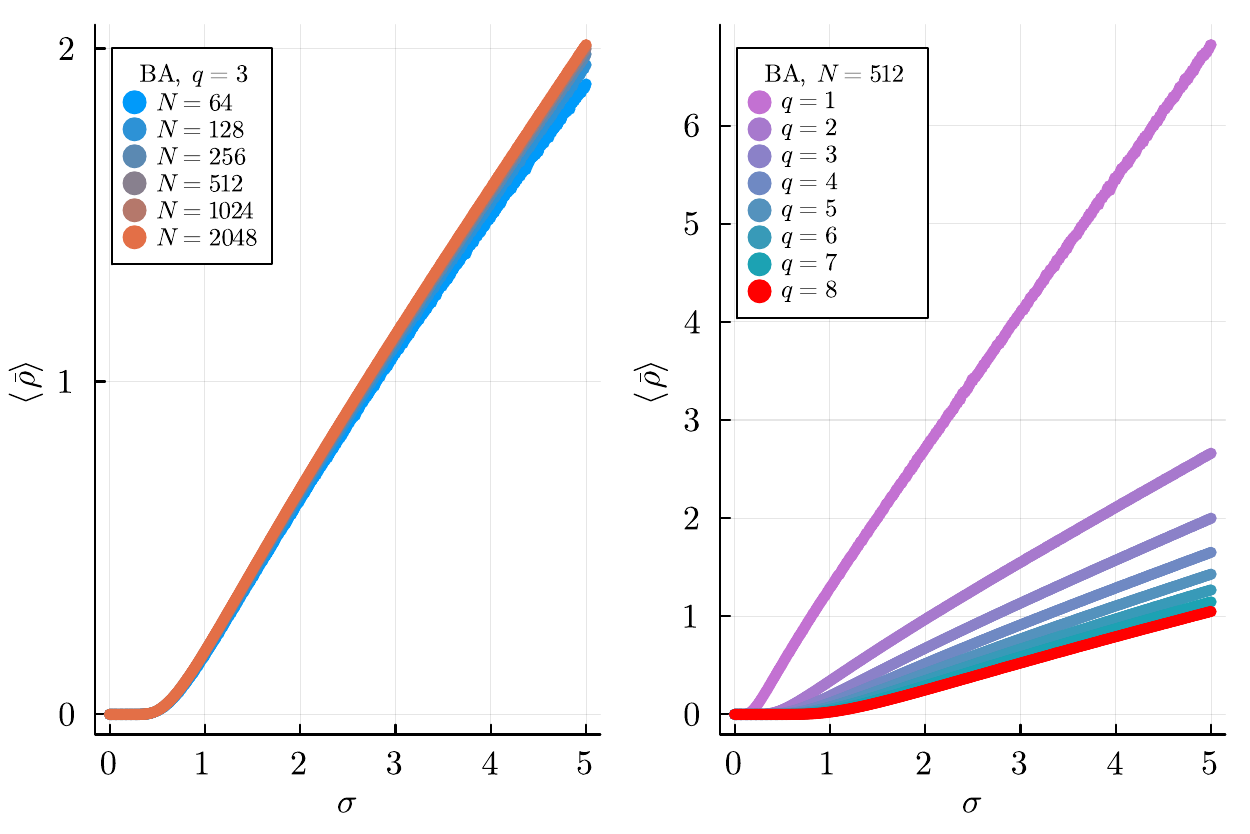}
\includegraphics*[scale=.4]{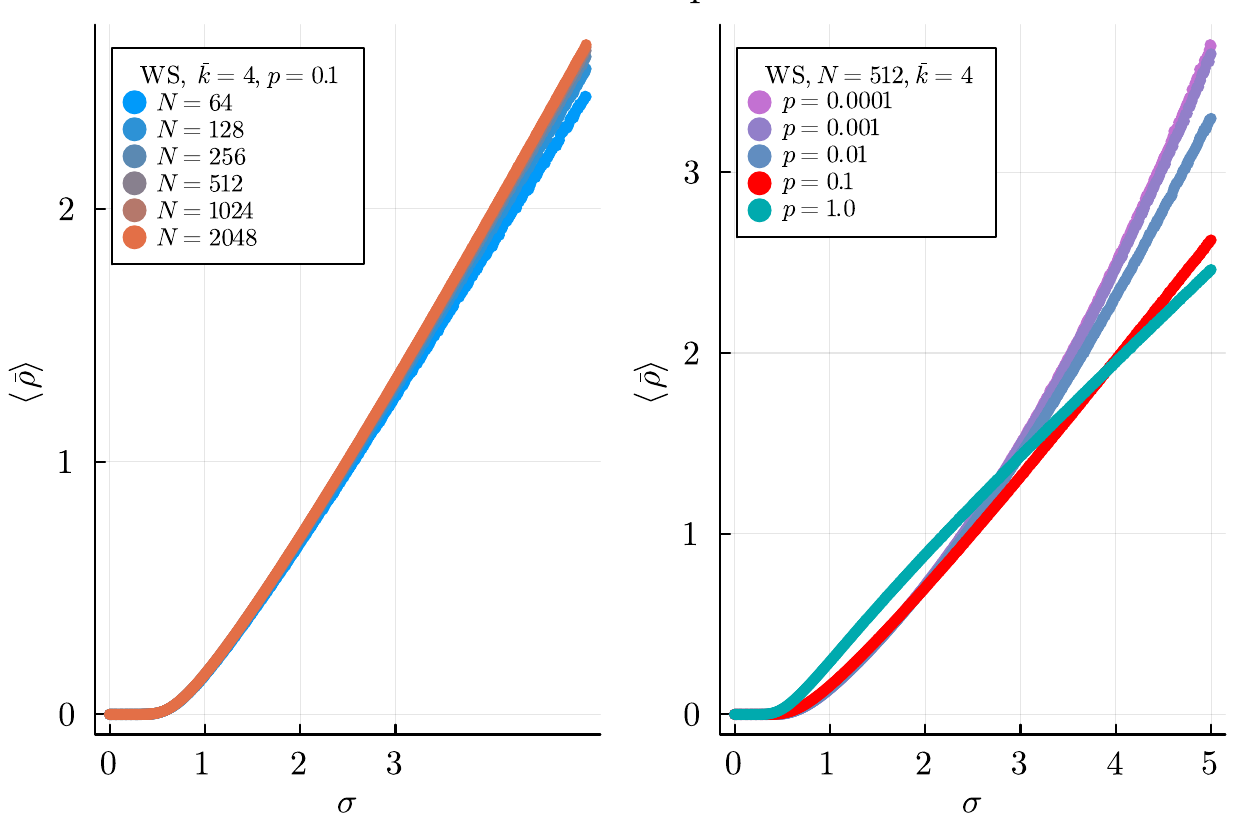}
\put(-483,140){\bf c)}
\put(-240,140){\bf d)}
\caption{
\label{fig:3}
(color online).
The mean of the average of the absolute differences of rankings $\avrg{\bar{\rho}}$ as a function of $\sigma$ for the same numerical experiments of Fig.~\ref{fig:2}.
The insets zoom in certain regions of the main plots.
The gray dashed lines in the insets represent guides to the eye.
}
\end{figure*}

To compare the true ratings $\hat{w}_i$ with the inferred ratings $w_i$, as well as the true rankings $\hat{r}_i$ with the inferred rankings $r_i$, the average absolute differences
$$
\tau
=
\frac{1}{N}
\sum_{i}
|\hat{w}_i-w_i|
$$
and
$$
\rho
=
\frac{1}{N}
\sum_{i}
|\hat{r}_i-r_i|
$$
are used, respectively.
These metrics are zero when the inferred ratings or rankings match the true ones, and greater than zero otherwise.
It is important to note that the means $\bar{\tau}(\sigma, a) = \int dw \, P(w | \sigma, a) \, \tau(w)$ and $\bar{\rho}(\sigma, a) = \int dw \, P(w | \sigma, a) \, \rho(w)$, while similar to order parameters, are more accurately described as {\em disorder parameters}, since they become different from zero when disorder kicks in. 
Here, the distribution $P(w | \sigma, a)$ represents the statistics of $w$ obtained from solving Eq.~\ref{eq:4} while sampling $\eta$ for a given network structure $a$.
In the experiments, the network-averaged quantities $\langle \bar{\tau} \rangle (\sigma) = \sum_a P(a | N, \theta) \bar{\tau}(\sigma, a)$ and $\langle \bar{\rho} \rangle (\sigma) = \sum_a P(a | N, \theta) \bar{\rho}(\sigma, a)$ are computed, where $N$ and $\theta$ represent relevant parameters of the network models. 
Specifically, $\theta$ denotes the coordination number $z$ for 1D lattices, the average degree $\bar{k}$ for ER networks, the number of new links $q$ per new node in BA networks, and the fraction $p$ of randomized links for WS networks with average degree $\bar{k}=4$.

Traditional solvers struggle to provide adequate solutions for Eqs.~\ref{eq:5} and~\ref{eq:6}. 
To address this, the recently introduced iterative solver \verb+MINRES-QLP+, developed by Choi {\em et al.}~\cite{choi2011minresqlp}, is employed to find the minimum norm solutions to the singular least-squares problems. 
This solver is available in the \verb+Krylov.jl+ package~\cite{montoison2023krylov} for the Julia programming language~\cite{bezanson2017julia}. 
A Jupyter notebook with the code used for the experiments can be found online~\cite{github_hon_ranking}.
The performance of \verb+MINRES-QLP+ was evaluated on nonperiodic 1D lattices of varying sizes $N$. 
For $\sigma = 0$, the solver demonstrates satisfactory numerical accuracy for network sizes up to approximately $N \approx 10^5$. 
However, to mitigate the impact of numerical errors, subsequent experiments are conducted on networks of considerably smaller sizes.

The following figures contain four panels, each corresponding to a different network model.
In each figure, panel (a) presents results for non-periodic 1D lattices, panel {\bf b)} for ER networks, panel (c) for BA networks and panel (d) for WS networks.
Furthermore, each panel is divided into two subpanels.
The left subpanel presents results for $\theta$ fixed and varying network sizes $N$, and the right subpanel for fixed $N$ and varying $\theta$.
The results are averaged over 5000 samples of both $a$ (the network structure) and $\eta$ (the disorder).
Where possible, a phenomenologically obtained approximate scaling relation for the studied quantities is indicated.

In Fig.~\ref{fig:2}, the mean of the average absolute differences, $\avrg{\bar{\tau}}$, between true and inferred ratings is plotted as a function of $\sigma$. 
For small $\sigma$, an approximately linear scaling of $\avrg{\bar{\tau}}$ with $\sigma$ is observed in most cases. 
However, deviations from this behavior emerge and are more pronounced in networks with smaller sizes $N$ or average degrees $\bar{k}$, especially in BA networks.
As shown in the left subpanels, $\avrg{\bar{\tau}}$ scales approximately as $\sim N^{1/2}$ for 1D lattices, while for all other network models, it appears largely independent of $N$. 
In the right subpanels, it is observed for 1D lattices that $\avrg{\bar{\tau}}$ scales as $\sim z^{-1.3}$ for sufficiently large values of $z$. 
Similarly, it scales as $\sim \bar{k}^{-2/3}$ for ER networks and $\sim q^{-2/3}$ for BA networks. 
In all cases, deviations from these scalings are observed when networks approach a sparse treelike structure.
For WS networks, $\avrg{\bar{\tau}}$ decreases with $\ln p$ in a sigmoidlike manner, as shown in the inset of the right subpanel of panel (d), where the transition point is found between the small-world and the lattice regimes.

In Fig.~\ref{fig:3}, the mean of the average absolute differences, $\avrg{\bar{\rho}}$, between true and inferred rankings is shown as a function of $\sigma$. 
Interestingly, as can be seen in more detail in the insets, two distinct regimes are observed.
For small values of $\sigma$, there is a region of perfect retrieval of the true ranking, where $\avrg{\bar{\rho}} = 0$. 
This behavior persists up to a non-trivial threshold, $\sigma^*$, after which $\avrg{\bar{\rho}}$ deviates from zero increasing monotonically with $\sigma$.
As seen in the right subpanels of panels (a), (b), and (c), a transition region emerges in networks with a sufficiently large connectivity, $\bar{k}$. 
In this region, $\avrg{\bar{\rho}}$ shifts from perfect retrieval to a regime where it grows approximately linearly with $\sigma$.
The distinction between these two regimes becomes more pronounced with increasing disorder in the network structure, as illustrated by the WS networks in the right subpanel of panel (d) as $p$ grows.

\begin{figure*}
\includegraphics*[scale=.4]{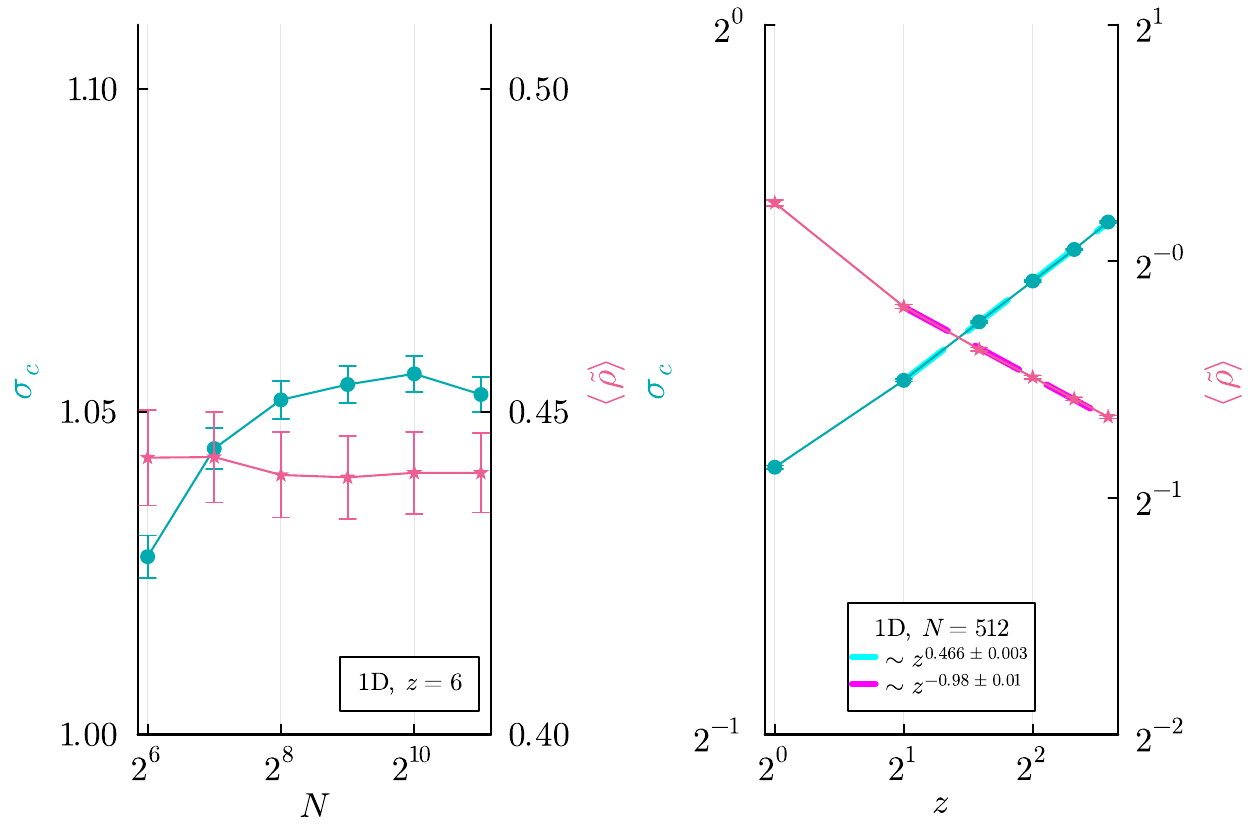}
\includegraphics*[scale=.4]{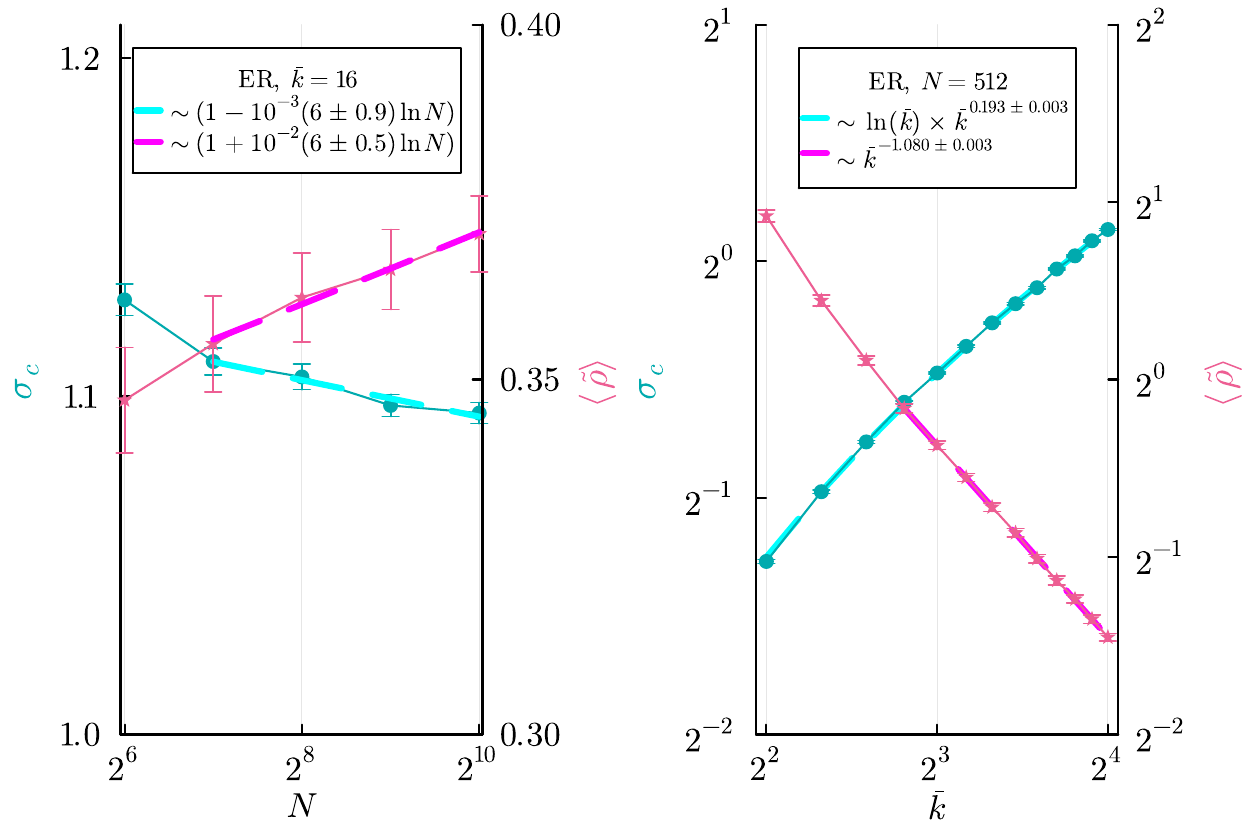}
\put(-483,140){\bf a)}
\put(-240,140){\bf b)}
\\
\includegraphics*[scale=.4]{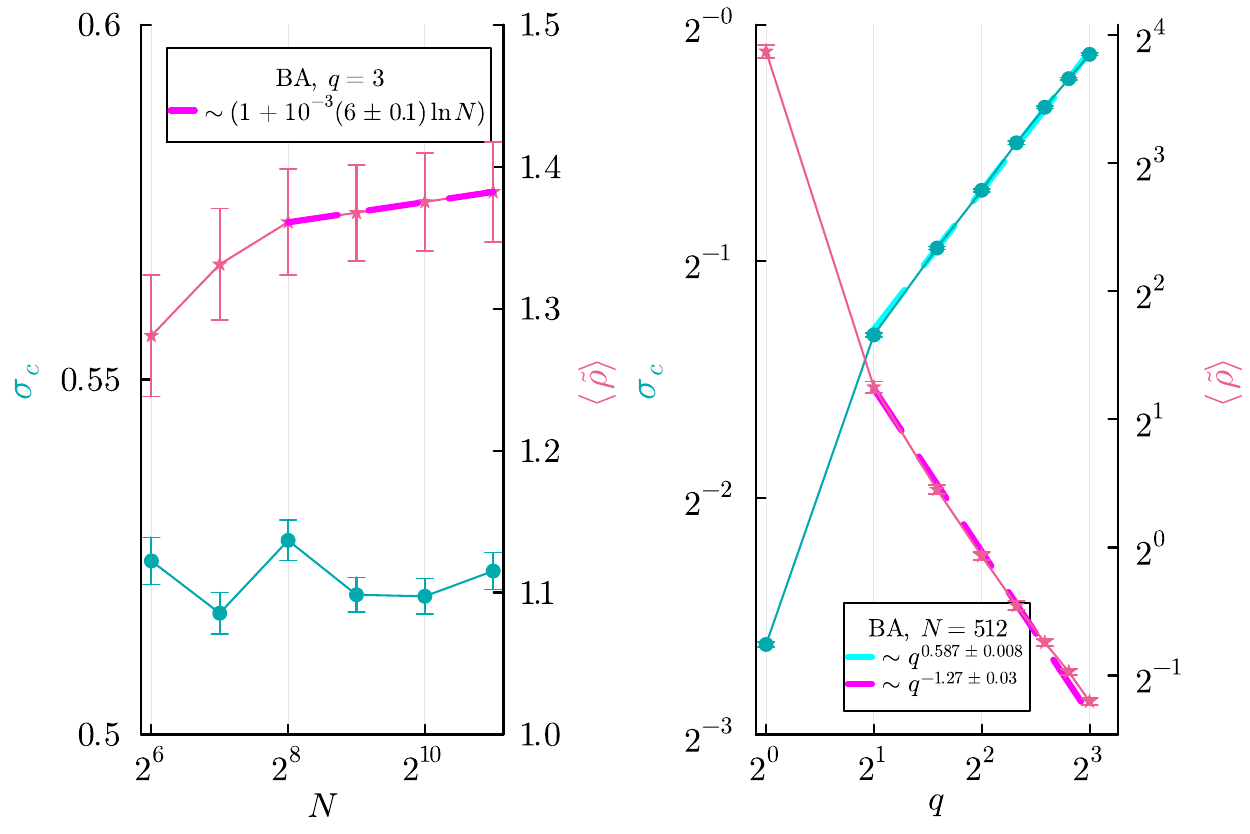}
\includegraphics*[scale=.4]{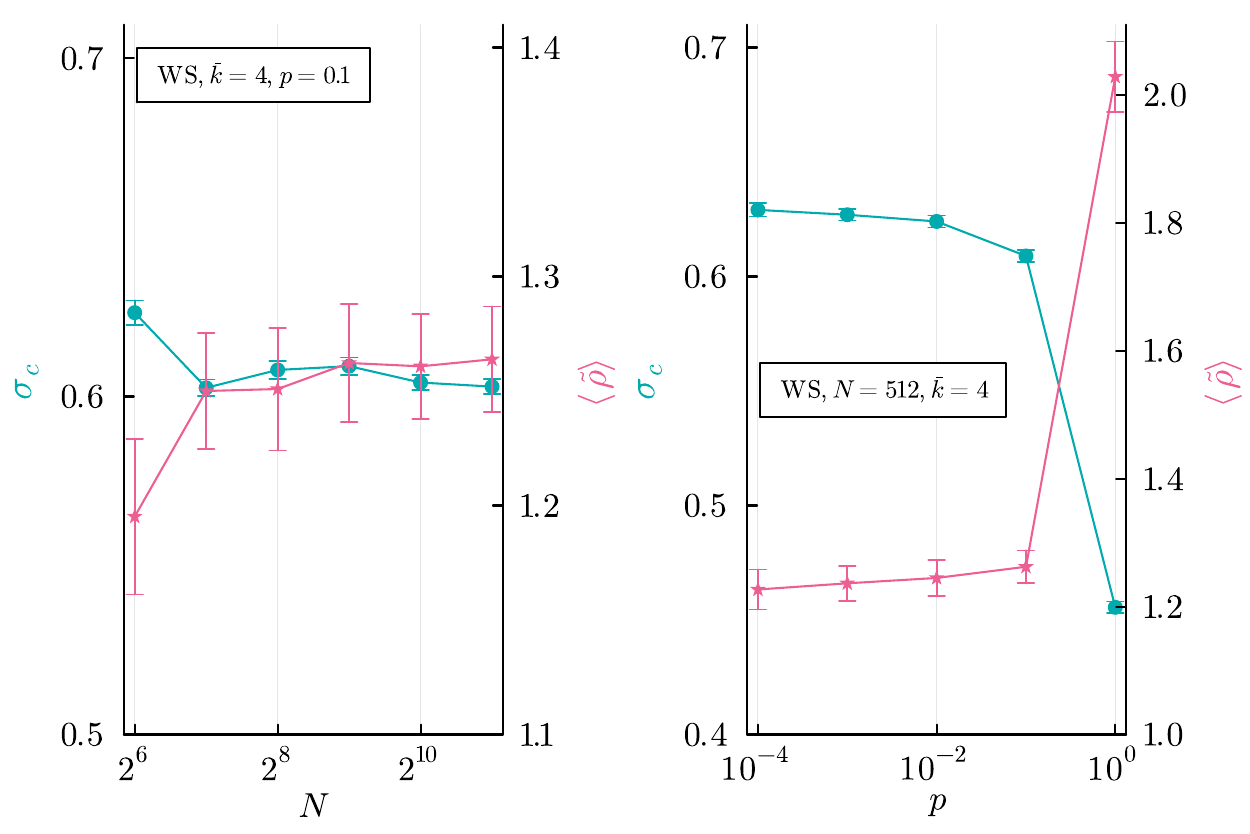}
\put(-483,140){\bf c)}
\put(-240,140){\bf d)}
\caption{
\label{fig:4}
(color online).
Transition points $\sigma_c$ (green circles) and peak heights $\langle \tilde{\rho} \rangle$ (pink stars) obtained from fits of the primitives of sigmoid functions to the curves $\langle \bar{\rho} \rangle$ vs $\sigma$ in Fig.~\ref{fig:3}.
Error bars are calculated through error propagation of the errors of the fitted parameters.
The dashed lines represent phenomenological scaling relations fitted to the data points.
}
\end{figure*}

To identify a transition point $\sigma_c$ within a transition region, the primitive
$$
\avrg{\bar{\rho}}(\sigma)
=
\frac{A}{B}
\ln
\big(
1+e^{B(\sigma-\sigma_c)}
\big)
$$
of the sigmoid function
$$
\avrg{\bar{\rho}}'(\sigma)
=
\frac{A}{1+e^{-B(\sigma-\sigma_c)}}
$$
is fitted to the curves of Fig.~\ref{fig:3}.
Here, $A$, $B$ and $\sigma_c$ are fitting parameters.
The primitive function approaches linear growth, $\avrg{\bar{\rho}}(\sigma) \approx A(\sigma - \sigma_c)$, for $\sigma \gg \sigma_c$, and approaches zero for $\sigma \ll \sigma_c$. 
The parameter $B^{-1}$ controls the width of the transition region.
The peak of the derivative of the sigmoid function
$$
\avrg{\bar{\rho}}''(\sigma)
=
\frac{ABe^{-B(\sigma-\sigma_c)}}{(1+e^{-B(\sigma-\sigma_c)})^2}
$$
determines the transition point $\sigma_c$. 
The range $[\sigma^*,\sigma^{**}]$ of fitted values for each curve is determined by maximizing the height of the peak, $\avrg{\tilde{\rho}}:=\avrg{\bar{\rho}}''(\sigma_c) = \frac{AB}{4}$, as a function $\sigma^{**}$.
Further details of the fitting procedure can be found in Figs.~1-8 and related text of the Supplemental Material (SM)~\cite{supplementalmaterial}.
The results of these fits are shown in Fig.~\ref{fig:4}, where the transition points $\sigma_c$ and peak heights $\avrg{\tilde{\rho}}$ are plotted as functions of the networks' parameters $N$ and $\theta$.
As seen in the left subpanels, the dependency of these quantities on the network size $N$ is negligible or relatively weak for all network models. 
In contrast, the right subpanels reveal different power-law scalings between these quantities and $\theta$ for the different network models and $N$ fixed.
More specifically, $\sigma_c$ grows with $z$, $\bar{k}$ and $q$ for 1D lattices, and ER and BA networks, respectively.
In particular, the scaling law has a logarithmic correction for ER networks.
On the other hand, the peak heigth $\avrg{\tilde{\rho}}$ decreases with $z$, $\bar{k}$ and $q$.
For WS networks, both $\sigma_c$ and $\avrg{\tilde{\rho}}$ remain approximately constant in the small-world regime, but undergo a sharp change once the lattice regime is reached.

A comprehensive summary of the scaling relations identified in this study is presented in Table I of the SM~\cite{supplementalmaterial}.

\section{
Discussion
\label{sec:discussion}
}

Interactions determine how entities behave in response to one another. 
While some interactions can be directly observed (like pairings between chess players), others have to be effectively inferred using response functions, correlation functions, perturbations or more sophisticate methods~\cite{barzel2013network}.
The pairwise comparisons of the present article define a standard network without higher-order interactions. 
However, the results of these comparisons introduce additional structure beyond the pairings themselves.
The HodgeRank method encodes this extra information into cochains by imposing the statistically parsimonious condition in Eq.~\ref{eq:4}.  
Minimizing $|f - d_0(w)|$ generally results in  a nonzero 1-cochain $s$, yielding higher-order simplicial interactions encoded in the exact 2-cochain $u$. 
While alternative minimization methods might overlook $u$, its presence remains implicit.  
More concretely, if agent $i$ defeats agent $j$, agent $j$ defeats agent $k$, and agent $k$ defeats agent $i$, then although the interactions are pairwise, their aggregate result encodes a cyclic flow of ratings represented by the higher-order interaction $u_{ijk}$. 
These interactions are not artifacts of the method.
Rather, HodgeRank follows the principle of parsimony, fully encoding available information in 0- and 1-cochains before assigning the remaining structure to 2-cochains.
In summary, while the network of pairwise comparisons does not exhibit higher-order topology, effective higher-order interactions emerge when aggregate results are incorporated into the data representation via HodgeRank.  

The scaling of $\langle \bar{\tau} \rangle$ versus $N$ in Fig.~\ref{fig:2} for the 1D lattice differs from that observed in other network models, except for networks of the WS model at $p\to 0$.  
This suggests that the presence of nonlocal cycles in the small-world regime influences the inference of ratings, a finding consistent with the behavior of $\langle \bar{\tau} \rangle$ as a function of $\ln p$ in the WS model.  
Regarding the scaling of $\langle \bar{\tau} \rangle$ with $z$, $\bar{k}$, and $q$, significant changes become evident as the networks approach a sparse, treelike regime, where the absence of redundant paths weakens the reinforcement of rating differences between arbitrary items.  
The transition from perfect retrieval ($\langle \bar{\rho} \rangle = 0$) to partial retrieval ($\langle \bar{\rho} \rangle > 0$) of the true rankings, as observed in Fig.~\ref{fig:3}, results from the progressive degradation of true ratings, which increases $\langle \bar{\tau} \rangle$, as the quenched disorder strength $\sigma$ increases.  
At low values of $\sigma$, the perturbations in ratings are too small to alter rankings, keeping $\langle \bar{\rho} \rangle$ at zero.  
As $\sigma$ surpasses $\sigma_c$, rating crossovers emerge, leading to ranking changes.
This effect is particularly evident in 1D lattices.

For sufficiently large connectivities, the transition region remains well defined and is approximately bounded by $[\sigma^*, \sigma^{**}]$, with the transition point $\sigma_c$ identified by the peak of the primitive of a sigmoid function.  
As connectivity decreases, or as the small-world property weakens with decreasing $p$, the transition region becomes less distinct, highlighting the critical role of redundant nonlocal paths in stabilizing rankings.  
These findings are further supported by the scaling behavior of $|f|$, $|s|$, $|h|$, and $|g|$ with respect to $\sigma$, across different network models, as shown for varying $N$ with fixed $\theta$ and varying $\theta$ with fixed $N$ in Figs.~9--12 of the SM~\cite{supplementalmaterial}, and the scaling relations for the average dimension $\langle \kappa_k \rangle$ of the vector spaces $C_k$ of Fig.~13 of the SM~\cite{supplementalmaterial}, which are also consistent with the observed behavior of $|s|$ and $|h|$.  
Moreover, this interpretation is reinforced by the approximately linear growth of $\langle \bar{\tau} \rangle$, $|h|$, and $|s|$ with $\sigma$, as shown in Fig.~\ref{fig:2} and Figs.~10~and~11 of the SM~\cite{supplementalmaterial}, respectively. 

In summary, the scaling relations found in this article demonstrate that the topology of the network of pairings significantly influences the inference of ratings and rankings, as well as its response to varying levels of quenched disorder.  
Specifically, while network size has little to no effect, increased connectivity, the presence of hubs, and the small-world property significantly enhance inference accuracy and delay the transition.  
Ultimately, these insights can help anticipate inference behavior and inform the design of better pairing strategies or rating methods in real-world scenarios, depending on the underlying network topology.  

\section{
Conclusions
\label{sec:conclusions}
}

This work explores the impact of complex network topologies on the performance of HodgeRank, a method developed to infer ratings from pairwise comparisons using a discrete or combinatorial formulation of differential geometry and Hodge theory over higher-order networks. 
The study presents a comprehensive scaling analysis of a variety of metrics, using well-established network models: 1D lattices, Erdős-Rényi random graphs, Barabási-Albert scale-free networks, and Watts-Strogatz small-world networks. 
Additionally, the analysis incorporates the effects of quenched disorder in the data, which negatively influences the inference of ratings and rankings.
The numerical experiments demonstrate that as the strength of disorder varies, a transition occurs from a regime of perfect retrieval to one of partial retrieval of the true rankings.
The scaling laws identified for the different studied quantities, highlight the critical role of the network structure.
This variability encompasses contributions from both the gradient and cyclic flows of the Hodge decomposition that form the foundation of the HodgeRank method.

Future research could employ traditional methods from statistical physics to further explore the properties of the observed transition and scaling relations.  
Additionally, this approach could be extended to the inference of ratings and rankings in empirical data, potentially leading to more realistic models of network topology and quenched disorder.  
In particular, it would be valuable to investigate how the transition is affected when a long-tailed distribution of quenched disorder is considered.  
Moreover, future studies could examine how quenched disorder influences the inferred ratings and rankings of individual nodes based on their local properties.  
Finally, leveraging the benchmark experiments developed in this work, it would be interesting to compare HodgeRank with alternative inference methods across a broader range of network models, particularly using modern computing architectures such as GPUs or analyzing very large datasets.  

\begin{acknowledgments}
The author acknowledges partial support from CONICET under grant PIP2021-2023 No. 11220200101100, and from SeCyT (Universidad Nacional de Córdoba, Argentina). Special thanks go to O.V. Billoni (FAMAF-UNC) for a careful reading of the article and S. Pinto (DF-UBA) for helpful discussions, and to the Centro de Cómputo de Alto Desempeño (CCAD) at the Universidad Nacional de Córdoba (UNC), which is part of SNCAD – MinCyT, Argentina, for providing computational resources.
\end{acknowledgments}



%

\appendix*

\section{\label{appA}
Discrete exterior calculus and combinatorial Hodge theory
}

This section provides an introductory review about discrete exterior calculus and combinatorial Hodge theory over simplicial complexes.
Where appropriate, analogies to the smooth standard version of differential geometry over manifolds are highlighted, but without forgetting that important differences exist between the discrete and the smooth settings~\cite{ribando2023combinatorial}.

\subsection{Notation}

If $\{e_i:i\in I\}$ is a basis of a vector space $V$ over a field $\FF$ and $I$ is a set of indices, then the linear decomposition of a vector $v\in V$ in this basis is denoted by $v=v^ie_i$ where Einstein's summation convention is assumed and $v^i\in \FF$.
Similarly, if $f\in V^*$ is a covector of the dual vector space of $V$, then $f=f_ie^i$ where $\{e^i:i\in I\}$ is the corresponding dual basis, $f_i\in \FF$ is a coordinate and $e^i(e_j)=\delta^i{}_j$ is Kronecker's delta.
If $a:V\to V$ is a linear operator, then there is a matrix of components $a_i{}^j$ such that $a(e_i)=a_i{}^je_j$.
In this way, $w^je_j=w=a(v)=v^ia(e_i)=v^ia_i{}^je_j$, so $w^j=v^ia_i{}^j=(a^T)^j{}_iv^i$ where $(a^T)^j{}_i$ are the components of the transpose of the matrix of components $a_i{}^j$~\footnote{The first index $i$ of the components $a_{ij}$, $a^{ij}$, $a_i{}^j$ and $a^i{}_j$ of a matrix indicates the row of the entry, while the second $j$ the column. 
There is no predefined relation between the left/right and the upper/lower properties of an index.}.
The kernel of $a$ is denoted by $\krn\,a=\{v\in V:a(v)=0\}$, the range is denoted by $\rng\,a=a(V)=\{a(v):v\in V\}$, the null is denoted by $\nll\,a=\dim\,\krn\,a$ and the rank is denoted by $\rnk\,a=\dim\,\rng\,a$.
An inner product over $V$ maps any two vectors $u,v\in V$ into a scalar $u\cdot v\in \FF$.
It is linear in the first argument if $\FF=\RR$ or antilinear if $\FF=\CC$, and always linear in the second argument.
Moreover, $v\cdot v\geq 0$ for all $v\in V$, so $|v|:=\sqrt{v\cdot v}$ is a norm over $V$ and, in particular, $v\cdot v=0$ implies $v=0$ whenever the inner product is non-degenerate.

\subsection{Simplicial complexes}

An abstract simplicial complex is a set $K$ of subsets of $\{0,1,\hdots,n\}$ that satisfies the property of topological closure: i.e. if $s\in K$ and $r\subset s$ is nonempty, then $r\in K$.
The elements $i$ of $\{0,1,\hdots,n\}$ are called nodes.
The elements of $K$ are called simplices.
For each simplex $s\in K$, there exist unique $i_0,\hdots,i_k$ in $\{0,1,\hdots,n\}$ such that $s=\{i_0,i_1,\hdots,i_k\}$ and $i_0<i_1<\hdots<i_k$.
Here, $\dim s=k=|s|-1$ is called the dimension of the simplex $s$ and $\dim K := \max_{s\in K} \dim\, s$ is the dimension of the complex $K$.
Correspondingly, $s$ is called a $k$-simplex and $K$ a $(\dim K)$-complex.
$K_k$ denotes the set of all $k$-simplices in $K$ and its cardinality is $\kappa_k:=|K_k|$, while $\kappa:=\sum_k\kappa_k=|K|$ is the cardinality of the complex $K$.
In particular, $K_{-1}=K_{n+1}=\emptyset$ are defined for convenience.
For any different nodes $i,j,k,l \in \{0,1,\hdots,n\}$, 0-simplices like $\{i\}$ are in one-to-one correspondence with nodes, 1-simplices like $\{i,j\}$ can be thought of links, 2-simplices like $\{i,j,k\}$ as triangles, 3-simplices like $\{i,j,k,l\}$ as tetrahedra, and so on.
In particular, a simplicial complex of dimension $\dim K=1$ is equivalent to a nondirected graph without self-links.

A simplex $r$ is a face of another $s$ if $r\subset s$.
There are ${k + 1 \choose q + 1}$ $q$-faces within a $k$-simplex.
Moreover, $r$ is a facet of $s$ if $r\subset s$ and $1+\dim\,r=\dim\,s$.
Correspondingly, $s$ is a coface/cofacet of $s$ if $r$ is a face/facet of $s$, respectively.
Additionally, $r$ and $s$ are incident to each other if $r$ is a facet of $s$ or $s$ is a facet of $r$.
Furthermore, $r$ and $s$ are lower adjacent to each other if they have a facet in common, and are upper adjacent to each other if they are both different facets of a common simplex.
In particular, two vertices of a common graph are upper adjacent if they are both different facets of a common edge.
The degree $k_s$ of a simplex $s$ is its number of cofacets.
The degree $k_s$ of a simplex $s$ should not be confused with the dimension $k$ of a $k$-simplex.
The generalized degree $\zeta_{sq}$ of a $k$-simplex $s$ is its number of $q$-cofaces.
In the particular case of common graphs or networks, $k_i=\zeta_{i1}$ for any node $i$.

A simplicial complex $K$ can be obtained from a graph $g$ where each $(k+1)$-clique of $g$ corresponds to a $k$-simplex of $K$.
Such simplicial complex obtained from $g$ is called the {\em clique complex} of $g$.
In particular, the $(k+1)$-clique complex of a graph $g$ is the simplicial complex obtained by considering only the $(q+1)$-cliques of $g$ for $q\leq k$.

\subsection{Chains}

The set of functions from $K$ to some number field $\FF$ is a vector space $C$ of vectors called {\em chains}.
The functions $e_s : K\to \FF$ for $s\in K$ such that $e_s(r)=\delta_{sr}$ form a canonical basis of $C$~\footnote{Here, $\delta_{sr}$ is the Kronecker delta.}.
In particular, if $s$ is a $k$-simplex, then $e_s$ is called a $k$-chain.
The subset of linear combinations of $k$-chains form a subspace $C_k$ of $C$.
Crucially,
$$
C=C_0\oplus C_1\oplus \hdots \oplus C_n
,
$$
where $C_0=C_n\cong\FF$, i.e. $C_0$ and $C_n$ are isomorphic to $\FF$.
For convenience, $C_{-1}=C_{n+1}=\{0\}$ are also defined.
Note, $\kappa_k=\dim\, C_k\leq {n\choose k}$ for $k=0,1,\ldots,n$, and $\dim\, C_{-1}=\dim\, C_{n+1}=0$. 
Therefore, $\dim\,C\leq \sum_{k=0}^n {n\choose k}=2^n$.
Using Einstein's notation, any chain of $C$ can be represented by a linear combination $c=c^se_s$ with unique coefficients $c^s\in \FF$ and where the summation index $s$ runs over $K$.

\subsection{Cochains}

The dual space $C^*$ of $C$ is called the vector space of {\em cochains} of $K$.
If $C_k^*$ is the dual space of $C_k$, then
$$
C^*=C_0^*\oplus C_1^*\oplus \ldots \oplus C_n^*
$$
and, for convenience, $C_{-1}^*=C_{n+1}^*=\{0\}$ are also defined.
The covectors in $C_k^*$ are called $k$-cochains.
In particular, $\{e^s:s\in K\}$ denotes the basis of $C^*$ dual to the canonical basis $\{e_s:s\in K\}$ of $C$.
As a consequence, $e^s(e_r)=\delta^s{}_r$ for all $s,r\in K$.
In this way, any cochain $f\in C^*$ can be represented by a linear combination $f=f_se^s$ of coefficients $f_s\in \mathbb{F}$.

The application of a cochain $f$ to a chain $c$ equals $f(c)=f_se^s(c^re_r)=f_sc^re^s(e_r)=f_sc^r\delta^s{}_r=f_sc^s\in \FF$.
In a sense, $f(c)$ is the discrete or combinatorial analog of the integration $\int_c f$ of form $f$ over a manifold $c$.
It should be remarked, however, that a simplicial complex is not always obtained from the discretization of a manifold.

\subsection{Orientation}

Let $e_{i_0i_1...i_k}:=e_{\{i_0,i_1,\ldots,i_k\}}$ when $0\leq i_0<i_1<...<i_k\leq n$.
Otherwise, let 
\begin{widetext}
$$
e_{j_0j_1\ldots j_k}
=
\left\{
\begin{array}{ll}
e_{i_0i_1\ldots i_k} & \mbox{if $j_0j_1\ldots j_k$ is an even permutation of $i_0i_1\ldots i_k$,} \\
-e_{i_0i_1\ldots i_k} & \mbox{if it is an odd permutation, and} \\
0 & \mbox{otheriwse.}
\end{array}
\right.
$$
\end{widetext}
The sign between $e_{j_0j_1\ldots j_k}$ and $e_{i_0i_1\ldots i_k}$ is called the orientation of $e_{j_0j_1\ldots j_k}$ relative to $e_{i_0i_1\ldots i_k}$.
Like the {\em orientation} of the axis of $\RR^n$, the orientation of the simplices is a matter of convention, like the arbitrariness in the enumeration of the nodes.

\subsection{
Boundaries
\label{sec:boundaries}
}

The boundary operator over chains $\partial : C\to C$ is defined by
$$
\partial = \partial_0+\partial_1+\ldots +\partial_n
,
$$
where $\partial_{-1}=\partial_0=\partial_{n+1}=0$, and
$$
\partial_k(e_{i_0i_1\ldots i_q})
=
\delta_{qk}
\sum_{j=0}^q
(-1)^j 
e_{i_0i_1\ldots \hat{i}_j\ldots i_k}
$$
for $0<k\leq n$, where the sign of $(-1)^j$ is called the orientation that the $k$-chain $e_{i_0i_1\ldots i_k}$ induces on the $(k-1)$-chain
$$
e_{i_0i_1\ldots \hat{i}_j\ldots i_k}:=e_{i_0i_1\ldots i_{j-1}i_{j+1}\ldots i_k}
\in C_{k-1}^*.
$$

The boundary operator $\partial$ satisfies the following crucial properties.
Firstly, it is a linear operator.
Secondly, $\partial_k(C)=\partial(C_k)\subseteq C_{k-1}$, meaning that the boundary of a simplex of dimension $k$ has dimension $k-1$.
Thirdly, $\partial^2=0$, meaning that a boundary has no boundary.
This third statement follows from
$$
\partial^2 
=
\sum_{jk}\,
\partial_j \partial_k
\nonumber
\\
=
\sum_{k}\,
\partial_{k}\partial_{k+1}
,
$$
and the fact that $\partial_{k}\partial_{k+1}=0$ is a consequence of the alternating definition of $\partial_k$ and that chains are oriented.

In matricial form 
$\partial_k(e_s)=(\alpha_k)_s{}^r e_r$.
In this way, $q^re_r=q=\partial_k(c)=\partial_k(c^se_s)
=c^s\partial_k(e_s)=c^s(\alpha_k)_s{}^re_r$.
Hence, $q^r=c^s(\alpha_k)_s{}^r=(\alpha_k^T)^r{}_sc^s$.
Note, $(\alpha_k)_s{}^r\neq 0$ if and only if $s$ and $r$ are incident to each other.
Hence, $\alpha_k\in \RR^{\kappa_{k}\times \kappa_{k-1}}$ or its transpose $\alpha_k^T$ are usually called the $k$th incidence matrix of the simplicial complex~\cite{grady2010discrete,newman2010networks,bianconi2021higher}.
The present work adopts the convention where the incidence matrix is identified with the transpose $\alpha_k^T$.

\subsection{Coboundaries}

The coboundary operator $d:C^*\to C^*$ is the discrete or combinatorial analog of the differential operator.
It is defined by
$$
[d(f)](c) = f[\partial(c)]   
$$
for all $f\in C^*$ and $c\in C$, and it can be recognized as the discrete or combinatorial analog of the generalized Stoke's theorem
$$
\int_c df = \int_{\partial c} f,
$$
which in turn is the generalization of the fundamental theorem of calculus to arbitrary dimensions.
Crucially, 
$$
d=d_0+d_1+\ldots +d_n,
$$
where $d_{-1}=d_n=d_{n+1}=0$,  $d_k(C^*)=d(C_k^*)\subseteq C_{k+1}^*$,
and $[d_k(f)](c)=f[\partial_{k+1}(c)]$ for all $c\in C$.
Moreover, $d^2=0$ and $d_k d_{k-1}=0$ easily follow from $\partial^2=0$.
This fact is the reason for which many differential equations in physics are of second order.

In matricial form 
$d_k(e^s)=(\beta_k)^s{}_r e^r$.
Hence, 
\begin{eqnarray}
[d_k(e^s)](e_r)
&=&
e^s[\partial_{k+1}(e_r)]
\nonumber
\\
\,[(\beta_k)^s{}_te^t](e_r)
&=&
e^s[(\alpha_{k+1})_r{}^le_l]
\nonumber
\\
(\beta_k)^s{}_t\,e^t(e_r)
&=&
(\alpha_{k+1})_r{}^l\,e^s(e_l)
\nonumber
\\
(\beta_k)^s{}_t\,\delta^t{}_r
&=&
(\alpha_{k+1})_r{}^l\,\delta^s{}_l
\nonumber
\\
(\beta_k)^s{}_r
&=&
(\alpha_{k+1})_r{}^s
.
\nonumber
\end{eqnarray}
In other words, $\beta_k=(\alpha_{k+1}^T)$.
Note, $g_re^r=g=d_k(f)=d_k(f_se^s)=f_sd_k(e^s)
=f_s(\beta_k)^s{}_re^r$.
Hence $g_r=f_s(\beta_k)^s{}_r=(\beta_k^T)_r{}^sf_s=(\alpha_{k+1})_r{}^sf_s$.

Note, $\grad = d_0:C_0^*\to C_1^*$ and 
$\curl = d_1:C_1^*\to C_2^*$ are recognized as discrete or combinatorial analogs of the gradient and the curl operators of standard differential calculus~\footnote{Strictly speaking, in standard differential calculus, the gradient operator is defined in a way that is metric dependent.}, respectively.
In particular, the known result $\curl\circ\grad=-d_1 d_0=0$ follows as a particular case of $d_k d_{k-1}=0$.
As an example on how these operators work, note that if $f\in C_0^*$ and $g=\grad\, f=d_0(f)$, then $g_{ij}=g(e_{ij})=[d_0(f)](e_{ij})=f[\partial_1(e_{ij})]=f(e_j-e_i)=f_j-f_i$.

\subsection{Inner products and geometry}

Consider a positive definite inner product from $C$ to $\RR$ denoted by $c\cdot q\in \RR$ for each $c,q\in C$.
By Riesz's representation theorem, a positive definite inner product over $C$ induces an isomorphism $C\ni c\to c^{\flat}\in C^*$ called {\em flat} defined by
$$
c^{\flat}(q) = c\cdot q
$$
for all $c,q\in C$.
Its inverse is denoted by $C^*\ni f\to f_{\sharp}\in C$ and is called {\em sharp}.
Together, $\flat$ and $\sharp$ are called {\em musical isomorphisms}.
They induce a positive definite inner product over $C^*$ given by $f\cdot g = f_{\sharp}\cdot g_{\sharp}$ for all $f,g\in C^*$.

Consider a non-degenerate inner product over $C$.
Its coordinates in the canonical basis are $m_{sr}=e_s\cdot e_r=e_r\cdot e_s=m_{rs}$ for $s,r\in K$.
Since $e_s^{\flat}\in C^*$, there exists $t_{sr}\in \RR$ such that $e_s^{\flat}=t_{sr}e^r$ and, therefore,
$$
m_{sq} = e_s\cdot e_q = e_s^{\flat}(e_q)=t_{sr}e^r(e_q)=t_{sr}\delta^r{}_q=t_{sq}.
$$
In consequence
\begin{eqnarray}
m_{sr}
&=&
e_s\cdot e_r=e_s^{\flat}\cdot e_r^{\flat}=(m_{st}e^t)\cdot (m_{rq}e^q) 
\notag
\\
&=&
m_{st}m_{rq}(e^t\cdot e^q)=m_{st}m_{rq}m^{tq}
\notag
\end{eqnarray}
so $0=m_{st}(\delta^t{}_r-m_{rq}m^{tq})$ and, as a result, $m_{st}m^{tr}=\delta_s{}^r$.
In other words, the matrices of components $m_{st}=e_s\cdot e_t$ and $m^{tr}=e^t\cdot e^r$ are the inverse of each other.
In the context of abstract simplicial complexes, the canonical basis $\{e_s : s \in K\}$ of $C$ is often assumed to be orthogonal, meaning that $m_{rs}$ and $m^{rs}$ represent components of diagonal matrices. 
In line with this convention, the present work adopts the trivial inner product $m_{rs} = \delta_{rs}$.

In differential geometry, an inner product on differential forms is typically defined using a volume form and an associated Hodge star operator, both of which depend on a metric tensor. When a simplicial complex arises from the discretization of a manifold, these structures may also be available, underscoring the close relationship between inner products on chains or cochains and the geometric properties of simplicial complexes. 
Regarding another subject, an hermitian inner product on $\mathbb{C}$ is preferred in some contexts. 
When this is the case, it can be incorporated into the formalism by complexifying $C$~\cite{roman2008advanced}.

\subsection{Dual boundaries}

The dual boundary operator $\partial^*:C\to C$ is defined as the inner-product adjoint of the boundary operator $\partial$, i.e. by
$$
\partial^*(c)\cdot q = c\cdot \partial(q)
$$
for all $c,q\in C$.
Crucially, $(\partial^*)^2=0$ follows from $\partial^2=0$, and
$$
\partial^*=\partial_0^*+\partial_1^*+\ldots +\partial_n^*,
$$
where  $\partial_{-1}^*=\partial_{0}^*=\partial_{n+1}^*=0$ and $\partial_k^*$ is the dual of $\partial_k$, so $\partial_k^*(C)=\partial^*(C_{k-1})\subseteq C_k$.

In matricial form 
$\partial_k^*(e_s) = (\mu_k)_s{}^r e_r$.
Hence,
\begin{eqnarray}
\partial_k^*(e_s)\cdot e_r
&=&
e_s\cdot \partial_k(e_r)
\nonumber
\\
(\mu_k)_s{}^t\, e_t\cdot e_r
&=&
(\alpha_k)_r{}^p\, e_s\cdot e_p
\nonumber
\\
(\mu_k)_s{}^t\, m_{tr}
&=&
(\alpha_k)_r{}^p\, m_{sp}
\nonumber
\\
(\mu_k)_s{}^t\, m_{tr}m^{rl}
&=&
(\alpha_k)_r{}^p\, m_{sp}m^{rl}
\nonumber
\\
(\mu_k)_s{}^t\, \delta_t{}^l
&=&
(\alpha_k)_r{}^p\, m_{sp}m^{rl}
\nonumber
\\
(\mu_k)_s{}^l
&=&
m^{lr}\,(\alpha_k)_r{}^p\, m_{ps}.
\nonumber
\end{eqnarray}
Moreover, in the particular case where $m_{sr}=\delta_{sr}$, so $m^{sr}=\delta^{sr}$, it is found that $(\mu_k)_s{}^t\delta_{tr}=(\alpha_k)_r{}^p\, \delta_{sp}$, so $(\mu_k)_s{}^r=(\alpha_k)_r{}^s$ and, therefore, $\mu_k=\alpha_k^T$.
Note, $q^re_r=q=\partial_k^*(c)=\partial_k^*(c^se_s)
=c^s\partial_k^*(e_s)=c^s(\mu_k)_s{}^re_r$.
Hence, $q^r=c^s(\mu_k)_s{}^r=(\mu_k^T)^r{}_sc^s=(\alpha_k)^r{}_sc^s$.

\subsection{Dual coboundaries}

The dual coboundary operator $d^*:C^*\to C^*$ is defined as the inner-product adjoint of the coboundary operator $d$, i.e. by
$$
d^*(f)\cdot g = f\cdot d(g)
$$
for all $f,g\in C^*$.
The dual coboundary operator is the discrete or combinatorial analog of the codifferential operator.
As before, $(d^*)^2=0$ follows from $d^2=0$, and
$$
d^*=d_0^*+d_1^*+\ldots +d_n^*,
$$
where $d_{-1}^*=d_{n}^*=d_{n+1}^*=0$ and $d_k^*$ is the dual of $d_k$, so $d_k^*(C^*)=d^*(C_{k+1}^*)\subseteq C_k^*$.

In matricial form
$d_k^*(e^s)=(\nu_k)^s{}_r e^r$.
Hence
\begin{eqnarray}
d_k^*(e^s)\cdot e^r
&=&
e^s\cdot d_k(e^r)
\nonumber
\\
(\nu_k)^s{}_t\,e^t\cdot e^r
&=&
(\beta_k)^r{}_p\,e^s\cdot e^p
\nonumber
\\
(\nu_k)^s{}_t\,m^{tr}
&=&
(\beta_k)^r{}_p\,m^{sp}
\nonumber
\\
(\nu_k)^s{}_t\,m^{tr}m_{rl}
&=&
(\beta_k)^r{}_p\,m^{sp}m_{rl}
\nonumber
\\
(\nu_k)^s{}_t\,\delta^t{}_l
&=&
(\beta_k)^r{}_p\,m^{sp}m_{rl}
\nonumber
\\
(\nu_k)^s{}_l
&=&
m_{lr}\,(\beta_k)^r{}_p\,m^{ps}
.
\nonumber
\end{eqnarray}
Moreover, in the particular case where $m^{sr}=\delta^{sr}$, so $m_{sr}=\delta_{sr}$, it is found that
$(\nu_k)^s{}_t\delta^{tr}=(\beta_k)^r{}_p\delta^{sp}$, so $(\nu_k)^s{}_r=(\beta_k)^r{}_s$ and, therefore, $\nu_k=\beta_k^T=\alpha_{k+1}$.
Observe that, $g_re^r=g=d_k^*(f)=d_k^*(f_se^s)
=f_sd_k^*(e^s)=f_s(\nu_k)^s{}_re^r$.
Hence, $g_r=f_s(\nu_k)^s{}_r=(\nu_k^T)_r{}^sf_s=(\beta_k)_r{}^sf_s=(\alpha_{k+1}^T)_r{}^sf_s$.

Note, $\dive = -d_0^*:C_1^*\to C_0^*$ can be recognized as the discrete or combinatorial analog of the divergence of standard differential calculus.

\subsection{Hodge decomposition}

Consider vector spaces $U$, $V$, and $W$ equipped with inner products, and let $a : V \to W$ and $b : U \to V$ be two linear operators such that
\begin{equation}
\label{eqA1}
ab=0.
\end{equation}
This implies that $\rng\,(b) \subseteq \krn\,(a)$, allowing the definition of the so-called (co)homology group~\footnote{Technically, this is a quotient space, but it can be considered an abelian group under addition.}
$$
H := \krn\,(a) / \rng\,(b) \cong \krn\,(a) \cap \krn\,(b^*) \cong \krn\,L,
$$
whose elements are referred to as (co)homology classes. Here, $L := a^*a + bb^* : V \to V$ is known as the Hodge Laplacian associated with $a$ and $b$.

It can be shown that
\begin{equation}
\label{eqA2}
V=\rng\,a^*\oplus\krn\,L\oplus\rng\,b
\end{equation}
which means that for any vector $v \in V$, there exists a unique decomposition
\begin{equation}
\label{eqA3}
v = a^*(w)+h+b(u)
\end{equation}
for some $w \in W$ and $u \in U$, where $h \in \krn\,L \subseteq V$ is the unique vector representing the (co)homology class $[v] \in \krn\,a / \rng\,b$ that satisfies $h \in (\rng\,b)^\perp = \krn\,b^*$. 
For this reason, $h$ is called the harmonic representative of $[v]$.

Together, Eqs.~\ref{eqA2}~and~\ref{eqA3} constitute the Hodge decomposition of $V$ and $v$, respectively. 
Notably, $\krn\,b^* = \rng\,a^* \oplus \krn\,L$ and $\krn\,a = \krn\,L \oplus \rng\,b$.
Moreover, $L$ is a positive semidefinite operator since $L = L^*$, $(a^*a)^* = a^*a$ and $(bb^*)^* = bb^*$.
Additionally, if $[X, Y] := XY - YX$, then $[L, a^*a] = [L, bb^*] = [a^*a, bb^*] = 0$.

\subsection{Homology}

In the context of simplicial complexes, homology can be derived through the application of Hodge decomposition, using $\partial_k$ and $\partial_{k+1}$ in place of operators $a$ and $b$, since $ab = \partial_k \partial_{k+1} = 0$.
As a result, the Hodge Laplacian takes the form
$$
\mathcal{L}_k = \mathcal{L}_k^{\downarrow} \oplus \mathcal{L}_k^{\uparrow} : C_k \to C_k,
$$
where
\begin{eqnarray}
\mathcal{L}_k^{\downarrow} &=& \partial_k^* \partial_k : C_k \to C_{k-1} \to C_k
\quad \text{and} \quad 
\notag \\
\mathcal{L}_k^{\uparrow} &=& \partial_{k+1} \partial_{k+1}^* : C_k \to C_{k+1} \to C_k.
\notag
\end{eqnarray}
This decomposition implies that
$$
C_k = \rng\,\partial_k^* \oplus \krn\,\mathcal{L}_k \oplus \rng\,\partial_{k+1},
$$
so for any $c \in C_k$, there exist unique components $p \in \rng\,\partial_k^* \subseteq C_k$, $h \in \krn\,\mathcal{L}_k \subseteq C_k$, and $q \in \rng\,\partial_{k+1} \subseteq C_k$ such that $c = p + h + q$.
Notably, $\krn\,\partial_{k+1}^* = \rng\,\partial_k^* \oplus \krn\,\mathcal{L}_k$ and $\krn\,\partial_k = \krn\,\mathcal{L}_k \oplus \rng\,\partial_{k+1}$.

The operators $\mathcal{L}_k$, $\mathcal{L}_k^{\downarrow}$, and $\mathcal{L}_k^{\uparrow}$ are often called the $k$th Laplacian, $k$th lower Laplacian, and $k$th upper Laplacian, respectively. 
These can be combined to define a Laplacian over $C$:
$$
\mathcal{L} = \partial^* \partial \oplus \partial \partial^* = \mathcal{L}_0 \oplus \cdots \oplus \mathcal{L}_n : C \to C.
$$
Notice, $\mathcal{L}_k^{\downarrow}$ and $\mathcal{L}_k^{\uparrow}$ combine information from $k$-chains with information from $(k-1)$-chains and $(k+1)$-chains, respectively.

The $k$th Dirac operator on $C$,
$$
\mathcal{D}_k = \partial_k^* + \partial_k,
$$
can also be introduced and is related to the concept of spinor~\cite{bianconi2021topological}. 
It satisfies $\mathcal{D}_k^2 = \mathcal{L}_k^{\downarrow} \oplus \mathcal{L}_{k-1}^{\uparrow}$, thereby combining information between $(k-1)$-chains and $k$-chains. 
Similarly, $\mathcal{D} = \partial^* + \partial = \mathcal{D}_0 + \cdots + \mathcal{D}_n$ satisfies $\mathcal{D}^2 = \mathcal{L}$.

The Betti numbers $\beta_k = \rnk\,H_k = \rnk\,\krn\,\partial_k - \rnk\,\rng\,\partial_{k+1}$ are topological invariants representing the ranks of the homology groups, which describe key properties of the underlying topological spaces. 
In the present context, where $C$ is a vector space, these homology groups are subspaces, so their ranks correspond to their dimensions.

Finally, if $m_{rs}$ represents the trivial metric $\delta_{rs}$, then $(\mathcal{L}_k^{\downarrow})^s{}_s = (\alpha_k)^s{}_r (\alpha_k^T)^r{}_s = k + 1$, which is the number of facets of $s$.
Similarly, $(\mathcal{L}_k^{\uparrow})^s{}_s = (\alpha_{k+1}^T)^s{}_r (\alpha_{k+1})^r{}_s = k_s = \zeta_{s,k+1}$, which is the number of cofacets or degree of $s$. 
For a general metric, $(\mathcal{L}_k^{\downarrow})^s{}_s$ and $(\mathcal{L}_k^{\uparrow})^s{}_s$ represent weighted versions of these quantities.

\subsection{Cohomology\label{appA:cohomology}}

Cohomology arises from the application of Hodge decomposition to cochains, using $d_k$ and $d_{k-1}$ in place of $a$ and $b$, respectively, such that $ab = d_k d_{k-1} = 0$.
Consequently, the Hodge Laplacian on cochains takes the form
$$
L_k = L_k^{\uparrow} \oplus L_k^{\downarrow} : C_k^* \to C_k^*,
$$
where
\begin{eqnarray}
L_k^{\uparrow} &=& d_k^* d_k : C_k^* \to C_{k+1}^* \to C_k^*
\quad \text{and} \quad
\notag \\
L_k^{\downarrow} &=& d_{k-1} d_{k-1}^* : C_k^* \to C_{k-1}^* \to C_k^*.
\notag
\end{eqnarray}
This decomposition implies that
$$
C_k^* = \rng\,d_k^* \oplus \krn\,L_k \oplus \rng\,d_{k-1},
$$
so for any $f \in C_k^*$, there exist unique components $s \in \rng\, d_k^* \subseteq C_k^*$, $h \in \krn\, L_k \subseteq C_k^*$, and $g \in \rng\, d_{k-1} \subseteq C_k^*$ such that $f = s + h + g$.
Typically, $s$, $h$, and $g$ are referred to as the coexact, harmonic, and exact components of $f$, respectively.
Additionally, it holds that $\krn\,d_{k-1}^* = \rng\,d_k^* \oplus \krn\,L_k$ and $\krn\,d_k = \krn\,L_k \oplus \rng\,d_{k-1}$.

The Laplacians $L_k^{\uparrow}$ and $L_k^{\downarrow}$ combine information from $k$-cochains with information from $(k+1)$-cochains and $(k-1)$-cochains, respectively.
Together, they define the Hodge Laplacian over $C^*$:
$$
L = d d^* \oplus d^* d = L_0 \oplus \dots \oplus L_n : C^* \to C^*.
$$
Moreover, in analogy with the Dirac operator $\mathcal{D}_k$ in homology, the $k$th Dirac operator on cochains,
$$
D_k = d_k^* + d_k,
$$
can also be defined.
It satisfies $D_k^2 = L_k^{\uparrow} \oplus L_{k+1}^{\downarrow}$, thus it combines information from $k$-cochains and $(k+1)$-cochains. 
Similarly, $D = d^* + d = D_0 + \dots + D_n$ satisfies $D^2 = L$.

The Hodge decomposition for cochains generalizes the Helmholtz decomposition from vector calculus in three dimensions to arbitrary discrete or combinatorial topologies and dimensions.
In this context, the coexact, harmonic, and exact components in the Hodge decomposition correspond to the solenoidal, harmonic, and gradient components of the Helmholtz decomposition.
Furthermore, $L_0 = d_0^* d_0 + d_{-1} d_{-1}^* = d_0^* d_0 + 0 = -\dive \circ \grad$ is known as the {\em graph Laplacian}, which serves as a discrete or combinatorial analog of the Laplace operator on scalar fields.
Similarly, $L_1 = d_0 d_0^* + d_1^* d_1 = \grad \circ \grad^* + \curl^* \circ \curl = -\grad \circ \dive + \curl^* \circ \curl$ is recognized as the {\em graph Helmholtzian}, which is a discrete or combinatorial analog of the vector Laplacian~\cite{lim2020hodge}.

When a simplicial complex represents a discretization of an oriented region in Euclidean space, a Hodge star operator can be defined to relate chains and cochains with volume forms.
Notably, in the case of a three dimensional triangulation, Hodge duality implies $\curl^* = \curl$, which is why there is no distinct name for $\curl^*$ in physics~\cite{grady2010discrete}.

\end{document}